\newif\ifcheckpagelimits
\checkpagelimitstrue
\checkpagelimitsfalse

\ifcheckpagelimits
 \documentclass[nofootinbib,prb,aps,twocolumn,showpacs,showkeys,%
 amsmath,amssymb,superscriptaddress,final,reprint,floatfix,longbibliography]{revtex4-1}
 \newcommand{\todo}[1]{}
\else
 \documentclass[pra,aps,twocolumn,showpacs,showkeys,%
 amsmath,amssymb,superscriptaddress,final,reprint,floatfix,longbibliography]{revtex4-1}
 \newcommand{\todo}[1]{{\pdfmargincomment[icon=Note,color=pink]{#1}}}
\fi

\usepackage{lineno}
  \usepackage{mathptmx}
  \usepackage{amssymb}
\usepackage{amsmath}
\usepackage{amsfonts}
\usepackage{subfigure}
\usepackage{dcolumn}
\usepackage{amsmath,amssymb}
\usepackage{bm}
\usepackage{color}
\usepackage{overpic}
\usepackage{latexsym}
\usepackage{epstopdf}
\usepackage{color}
\usepackage[english]{babel}
\usepackage{latexsym}
\usepackage{stmaryrd}

\usepackage{braket}

\definecolor{mygrey}{gray}{0.35}
\definecolor{myblue}{rgb}{0.2,0.2,0.8}
\definecolor{myzard}{cmyk}{0,0,0.05,0}
\definecolor{mywhite}{rgb}{1,1,1}
\definecolor{myred}{rgb}{1,0.,0.3}

\usepackage[colorlinks=true,citecolor=myblue,linkcolor=myred]{hyperref}
\DeclareMathAlphabet{\mathpzc}{OT1}{pzc}{m}{it}

 \def\ee{\mathord{\rm e}}
 
 \def\ii{\mathord{\rm i}}

\def\half{\textstyle\frac{1}{2}}

\def\fourth{\textstyle\frac{1}{4}}

\renewcommand{\ii}{{\rm i}}
\renewcommand{\ee}{{\rm e}}
\def\beq{\begin{equation}}
\def\eeq{\end{equation}}

\def\barray{\begin{eqnarray}}
\def\earray{\end{eqnarray}}

\begin{document}

\title{$\mathbb{Z}_N$  gauge theories coupled to topological fermions: QED$_2$ with a quantum-mechanical $\theta$ angle}

%\title{$\mathbb{Z}_N$ lattice gauge theories coupled to topological fermions: classical and quantum simulations}

\author{G. Magnifico}
\affiliation{Dipartimento di Fisica e Astronomia dell'Universit\`a di Bologna, I-40127 Bologna, Italy}
\affiliation{INFN, Sezione di Bologna, I-40127 Bologna, Italy}

\author{D. Vodola}
\affiliation{Department of Physics, College of Science, Swansea University, Singleton Park, Swansea SA2 8PP, United Kingdom}

\author{E. Ercolessi}
\affiliation{Dipartimento di Fisica e Astronomia dell'Universit\`a di Bologna, I-40127 Bologna, Italy}
\affiliation{INFN, Sezione di Bologna, I-40127 Bologna, Italy}

\author{S. P. Kumar}
\affiliation{Department of Physics, College of Science, Swansea University, Singleton Park, Swansea SA2 8PP, United Kingdom}

\author{M. M\"{u}ller}
\affiliation{Department of Physics, College of Science, Swansea University, Singleton Park, Swansea SA2 8PP, United Kingdom}

\author{A. Bermudez}
\affiliation{Departamento de F\'isica Te\'orica, Universidad Complutense, 28040 Madrid, Spain}
%that for different values of $N$
% by performing a scaling analysis of two order parameters and of the block entanglement entropy we provide accurate estimates of the critical lines, and their universality classes. Moreover,
\begin{abstract}
We present a detailed study of the topological Schwinger model [\href{http://dx.doi.org/10.1103/PhysRevD.99.014503}{Phys. Rev. D {\bf 99,} 014503 (2019)}],  which describes (1+1) quantum electrodynamics of an Abelian  $U(1)$ gauge field coupled to a symmetry-protected topological matter sector,  by means of a class of $\mathbb{Z}_N$ lattice gauge theories. Employing  density-matrix renormalization group techniques that exactly implement Gauss' law, we show that these models host a correlated topological phase for different values of  $N$, where fermion correlations arise through inter-particle interactions mediated by the gauge field. Moreover, by a careful finite-size scaling,   we show that  this phase is stable in the large-$N$ limit, and that the phase boundaries are  in accordance to bosonization predictions of the  $U(1)$  topological Schwinger model. Our results demonstrate that $\mathbb{Z}_N$ finite-dimensional gauge groups offer a practical route for an efficient classical simulation of equilibrium properties of   electromagnetism with topological  fermions. Additionally, we describe a scheme for the quantum simulation of a topological Schwinger model  exploiting spin-changing collisions in boson-fermion mixtures of ultra-cold atoms in optical lattices. Although technically challenging, this quantum simulation  would provide an alternative to classical density-matrix renormalization group techniques, providing also an efficient route to explore real-time non-equilibrium phenomena.
\end{abstract}

%\pacs{TBD}
\ifcheckpagelimits\else
\maketitle
\fi
\setcounter{tocdepth}{2}
\begingroup
\hypersetup{linkcolor=black}
%\tableofcontents
\endgroup

 \ifcheckpagelimits
\else

\maketitle
\fi
\setcounter{tocdepth}{2}
\begingroup
\hypersetup{linkcolor=black}
%\tableofcontents
\endgroup

\section{\bf Introduction}
\label{sec:introduction}

One of the greatest achievements of condensed-matter physics in the last century has been the classification of states of matter by the phenomenon of spontaneous symmetry breaking~\cite{landau_sb}. This has lead to a universal description of a wide variety of quantum states of matter, and phase transitions thereof, through the identification of effective field theories involving local order parameters. However, after the discovery of integer quantum Hall states~\cite{iqhe_exp,qhe_invariant}, it was soon realized that important quantum phases that are not contained in the Ginzburg-Landau symmetry breaking paradigm can exist. In particular, the description of these new states of matter requires the introduction of non-local order parameters, and  the interplay of certain mathematical tools of topology, such as topological invariants, with global protecting symmetries of the microscopic models. Two  states that show different topological invariants cannot be adiabatically connected, even if they   share the same symmetry. Instead,   quantum phase transitions induced by symmetry-preserving couplings  must occur, which cannot be accounted for by the symmetry-breaking principle. These so-called symmetry protected topological (SPT) phases lie at the focus of current research in condensed matter, and have interesting connections to high-energy physics. Since the pioneering work of F.D.M. Haldane~\cite{haldane_spt},  C. Kane and E. Mele~\cite{kane_mele_z2}, a variety of SPT phases beyond the integer quantum Hall states have been identified. These phases arise in different symmetry classes and dimensionalities, such as the so-called topological insulators and superconductors~\cite{kane_rmp_ti, qi_rmp_ti}, and some of them have already been realized experimentally~\cite{tis_exp}.

In analogy to the  quantum Hall effect, where the introduction of inter-particle interactions can lead to strongly-correlated phases with exotic properties~\cite{fqhe_exp} (e.g. excitations with fractional statistics~\cite{fqhe_laughlin}), a problem of current interest in the community is to understand the fate of this variety of SPT phases in the presence of interactions~\cite{corr_ti}. The typical models considered so far have mostly focused on instantaneous interactions of various ranges (e.g. screened Coulomb interactions, or truncated versions thereof, such as Hubbard-type couplings). More recently, some studies~\cite{spt_gauge, spt_schwinger,gonzalez_cuadra_18,gonzalez_cuadra_19} have started to explore strongly-correlated SPT phases with inter-particle interactions that are carried  by auxiliary fields  locally coupled  to the matter sector, avoiding in this way the  action at a distance of effective models with instantaneous interactions. This leads to interesting scenarios with intertwined topological phases that simultaneously display  symmetry breaking, as characterised by a local order parameter, and  topological symmetry protection, as described by a non-zero topological invariant~\cite{gonzalez_cuadra_18,gonzalez_cuadra_19}.  This  more general framework of interacting  SPT phases also allows one to explore models where interparticle interactions are carried by gauge bosons, and constrained by imposing invariance under local gauge symmetries~\cite{spt_gauge, spt_schwinger}. These microscopic models parallel  the use of  gauge theories in the fundamental interactions of particle physics~\cite{YM_non_abelian,qft_book}, and are particularly interesting in the context of non-perturbative effects in high-energy physics~\cite{qcd_confinement}. 

As outlined above, the study of SPT phases in condensed matter originated with the study of two- and three-dimensional topological insulators~\cite{kane_rmp_ti, qi_rmp_ti}. These phases of matter can be understood as lower-dimensional counterparts of the so-called domain-wall fermions in lattice gauge theories~\cite{dw_kaplan,kaplan_review,dw_fermions}.
In this context~\cite{gattringer_lang_book,kogut_review}, non-perturbative aspects of chiral gauge theories are studied on the boundary of a (4+1)-dimensional lattice, such that the existance of chiral fermions can be connected to a non-zero topological invariant and the so-called topological edge states. In the domain-wall fermion approach to LGTs~\cite{gattringer_lang_book}, one introduces an auxiliary synthetic dimension, such that Dirac fermions with a well-defined chirality become exponentially localised within  the boundaries of this auxiliary dimension, and interact via local couplings to the Abelian or non-Abelian gauge fields. We note that one is interested only in the physics that  takes place  on such boundaries, and not along the auxiliary dimension, and may decide to place the gauge fields only along the boundary. In the study of correlated SPT phases interacting via gauge bosons, the auxiliary dimension represents the physical bulk of the material. Therefore, gauge fields must thus also be defined within the bulk, and be minimally coupled to the bulk matter according to gauge-symmetry constraints. This difference becomes  relevant for the (1+1) model studied in this work, as one can define a gauge field theory in the 1D bulk, but the zero-dimensional edges cannot encompass a quantum field theory with gauge symmetry. 

A  well-studied gauge theory in low dimensions is the so-called Schwinger model~\cite{schwinger_model}, which describes quantum electrodynamics of relativistic fermions on the line. The Schwinger model can be considered as a paradigmatic toy model that unveils certain fundamental aspects that also play a role in higher-dimensional gauge theories, as well as interesting aspects that arise only as a consequence of the low dimensionality. One finds several peculiarities already  at the classical level,  since magnetic fields and photons do not exist, and static charges generate electric fields that do not decay with the distance. Upon quantization, one encounters the phenomenon of chiral symmetry breaking~ as the massless fermions pair and only massive excitations are found~\cite{schwinger_model}. This model has also played a key role in our current understanding of anomalies~\cite{anomalies_schwinger}, and the importance of the so-called topological  angles in the degeneracy of  vacua~\cite{theta_angle_schwinger}. In this manuscript, we shall be concerned with the massive Schwinger model, which describes massive relativistic fermions coupled via the one-dimensional electromagnetic field~\cite{massive_schwinger_shielding,massive_schwinger_theta}, and its various lattice discretizations.

 More specifically, in Ref.~\cite{spt_schwinger}, we introduced an alternative discretization of the continuum Schwinger model~\cite{schwinger_model} leading to the {\it topological Schwinger model}: an Abelian  gauge theory that regularizes quantum electrodynamics in (1+1) dimensions (QED$_2$), and describes the coupling of the electric field to a fermionic SPT matter sector. In contrast to the standard discretization of the massive Dirac fields, where one explicitly breaks translational invariance by using a  staggered mass~\cite{ham_lgt}, we chose to break the symmetry by a dimerized  tunnelling. In the continuum limit, such a tunnelling   leads to Dirac fermions with a topological mass, and a  matter sector that can be described as a fermionic SPT phase with a non-vanishing topological invariant and localised edge states~\cite{spt_schwinger}. Using bosonization, we showed that the coupling of this topological matter sector to the Abelian gauge field generates a quantum-mechanical vacuum $\hat{\theta}$ angle, which is no longer fixed by an external background electric field~\cite{massive_schwinger_shielding,massive_schwinger_theta}, but depends on the fermionic density on the edge states. We also used the bosonized theory to predict fundamental differences between the standard Schwinger model with a background  $\theta=\pi$ angle, and this  {topological Schwinger model}  with a quantum-mechanical $\hat{\theta}$ angle~\cite{spt_schwinger}. Let us note that, in the context of (3+1) time-reversal topological insulators, the response of the material to external electromagnetic fields can be described by an effective gauge field theory with an adiabatic axion field $\theta(\boldsymbol{x},t)$~\cite{axion_ti}, which becomes a vacuum $\theta$ angle when the axion field is homogeneous and constant, and is related to the topological invariant characterising the matter sector~\cite{kane_rmp_ti, qi_rmp_ti}. Our bosonization results can be understood as a lower-dimensional counterpart of the axion vacuum angle, which neatly captures the bulk-edge correspondence. Instead of relating  the vacuum $\theta$ angle to a topological invariant~\cite{axion_ti}, the bosonization approach upgrades this angle to a  quantum-mechanical operator $\hat{\theta}$ that depends on the topological edge-state operators, and may have its own dynamics beyond the adiabatic regime.

 In order to test these bosonization predictions, we performed  a numerical study based on the density-matrix renormalization group (DMRG)~\cite{spt_schwinger}. Instead of exploring the compact $U(1)$ LGT, one may employ the finite-dimensional  gauge groups $\mathbb{Z}_N$ for different values of $N$~\cite{hamiltonian_z_n_gauge,zn_presentation,alternative_z_n_gauge,alternative_z_n_gauge,zn_study}.  In Ref.~\cite{spt_schwinger}, we explored in detail the $\mathbb{Z}_3$ LGT with a topological matter sector,  showing that the above bosonization predictions qualitatively capture the physics already at $N=3$. We found that the phase diagram, as a function of the tunnelling dimerization and the gauge coupling, contains a wide region with a correlated SPT phase, where the correlations are induced by the fermion-fermion interactions mediated by the gauge field. This SPT phase is surrounded by other phases that also appear in the standard  Schwinger model at $\theta=\pi$, namely a fermion condensate and a confined phase~\cite{massive_schwinger_shielding,massive_schwinger_theta}. We found no adiabatic path connecting these phases, but instead  Ising-type quantum phase transitions take place. In this work, we perform a quantitative benchmark of  the  bosonization predictions by exploring larger values of $N$, and performing a  scaling analysis to extract  large-$N$ limit~\cite{zn_study}, where one expects to recover the predictions for the compact $U(1)$ LGT.  We provide a more detailed description of the numerical DMRG simulations presented in~\cite{spt_schwinger},  and  present new numerical results that provide further information about the properties of the topological Schwinger model. In particular, we quantitatively confirm  the bosonization predictions about the precise location of the  critical lines at weak couplings.  This quantitative analysis proves that the $\mathbb{Z}_N$ finite-dimensional gauge groups offer a practical route for an efficient classical simulation of equilibrium properties of   QED$_2$ with topological  fermions. 
 
Let us  note that LGTs are not mere computational tools, but may also become  realized  in experimental systems far from the high-energy-physics domain. In particular, some of the discretized LGTs  could be implemented with highly-tunable quantum systems ~\cite{QS_cold_atoms,QS_trapped_ions} in the low-energy domain: quantum simulators (QSs)~\cite{feynman_qs,qs_goals}. These QSs have the potential of becoming an alternative to Monte-Carlo-based simulation of LGTs, addressing  non-perturbative questions about the model  using quantum-mechanical hardware, and evading in this way current numerical limitations regarding real-time observables, or the occurrence of the fermion sign problem. In Ref.~\cite{spt_schwinger}, we outlined that it should be possible to implement the aforementioned topological Schwinger model using ultra-cold atoms in optical lattices.  Building on previous proposals for the quantum simulation of LGTs~\cite{qs_LGT_review},  we describe in this manuscript a concrete and detailed proposal  for the quantum simulation of this type of LGTs coupled to a topological matter sector. In particular, we show how to combine  spin-changing collisions in boson-fermion mixtures of ultra-cold atoms, and Floquet engineering by periodic drivings, in order to realize the desired topological Schwinger model.

This paper is stuctured as follows. In Sec.~\ref{sec:topological_schwinger_model}, we start by reviewing the  properties of standard Schwinger model in the continuum,  and describe its  discrete lattice version through the  Kogut-Susskind Hamiltonian approach employing staggered fermions~\cite{ham_lgt}. This discussion allows us to introduce in a natural way the alternative discretization that will give rise to a topological matter sector, and the use of $\mathbb{Z}_N$ gauge groups as a proxy of the Abelian $U(1)$ gauge fields.  In Sec.~\ref{sec:dmrg}, we start by reviewing some results obtained in Ref.\cite{spt_schwinger} for the $\mathbb{Z}_3$ LGT coupled to a topological matter sector, and describe the corresponding phase diagram. Then, by using  DMRG, we present new results on the scaling of the block entanglement entropies, providing accurate estimates of the universality classes of the critical lines. We extract the central charges of the underlying conformal field theories for the critical lines, and show that the $c = 1$ massless Dirac fermion of the non-interacting model splits into a couple of $c = 1/2$ massless Majorana fermions as soon as the gauge coupling $g$ is switched on. Moreover, we consider the cases when the gauge fields belong to the groups $\mathbb{Z}_N$ with $N=5$ and $N=7 $, performing a careful finite-size scaling in the thermodynamic limit of two order parameters for extracting the position  of the critical lines and their universality classes. In addition, we study the  scaling with $N\to\infty$ of the critical points for accessing the $U(1)$ limit of the $\mathbb{Z}_N$ topological Schwinger model, and compare the numerical large-$N$ results with the bosonization predictions for the $U(1)$ LGT. In Sec.~\ref{sec:cold_atoms_qs}, we introduce a scheme based on spin-changing collisions and periodic lattice modulations, sometimes referred to as Floquet engineering~\cite{floquet_cold_atoms}, for the realization of the topological Schwinger model in a Bose-Fermi mixture of ultra-cold atoms in a 1D optical lattice. This scheme provides a promising, albeit challenging, route for a future experimental quantum simulation. Finally, Sec.~\ref{sec:concl} contains concluding remarks and gives outlook on future work.

\section{\bf The topological Schwinger model}
\label{sec:topological_schwinger_model}

\subsection{Continuum Schwinger model with a  $\theta$ angle }

In this section, we start by reviewing the continuum massive Schwinger model \cite{massive_schwinger_shielding, massive_schwinger_theta},  which describes the interaction of a massive Dirac fermion interacting with the electromagnetic field. In a (1+1)-dimensional Minkowski spacetime with coordinates $x^\mu$,  $\mu\in\{0,1\}$ (i.e. $x=(t,{\rm x})$), and  after setting  $\hbar=c=1$, the Lagrangian density  that dictates the dynamics of the fermionic and gauge fields is given by 
\beq
\label{eq:schwinger_lagrangian}
\mathcal{L}_{{\rm S}}(m,g)=\overline{\Psi}(\ii\gamma^\mu(\partial_\mu+\ii gA_\mu)-m){\Psi}-\fourth F^{\mu\nu}F_{\mu\nu},
\eeq
  where $\partial_\mu=\partial/\partial x^{\mu}$,  $A_\mu(x)=\eta_{\mu\nu}A^\nu(x)$, and we use the repeated-indexes summation criterion with Minkowski's metric $\eta={\rm diag}(1,-1)$. In the expression above, we have introduced the Dirac matrices satisfying the anti-commutation relations $\{\gamma^\mu,\gamma^\nu\}=2\eta^{\mu\nu}$. In  (1+1) dimensions, the Dirac matrices  can be represented in terms of  Pauli matrices, and our choice is $\gamma^0=\sigma^y$, and $\gamma^1=\ii\sigma^z$. We have also introduced  $\overline{\Psi}(x)=\Psi^{\dagger}(x) \gamma^0$,  and the (bare) coupling  $g$ of the fermion current to the gauge field with  a Faraday  tensor $F_{\mu\nu}=\partial_{\mu}A_{\nu}-\partial_{\nu}A_{\mu}$. With this notation, the fields have the classical mass (energy) dimensions $d_{\psi}=1/2$ and $d_{A_\mu}=0$, while the mass and gauge coupling have $d_m=d_g=1$.
  
The Schwinger model is the simplest tractable QFT that captures some of the most significant non-perturbative effects  
displayed by non-Abelian gauge theories in higher dimensions. In the massless limit  $m=0$, it was solved exactly by J. Schwinger~\cite{schwinger_model}, who showed that the spectrum can be described by non-interacting bosons with a mass proportional to the coupling strength. In this massless limit, single-fermion excitations do not appear in the spectrum,  but only massive neutral quasiparticles composed of bound fermion-antifermion pairs. This is known as fermion trapping in the high-energy context~\cite{massive_schwinger_shielding}, and it is reminiscent of exciton formation in condensed matter~\cite{exciton_formation_theory}. In this later context, electrons and holes are attracted by the electromagnetic forces, forming bound states  and leading a neutral bosonic quasi-particle.  There are, however,  also clear differences due to the relativistic nature of the fermions and the peculiarities of the Coulomb  force in such reduced dimensions.

The Schwinger model also provides a neat framework for understanding  important properties that are absent in classical electrodinamics, such as the chiral anomaly~\cite{schwinger_anomaly}, or  the so-called vacuum $\theta$ angle, which modifies the Lagrangian as
\beq
\label{eq:schwinger_lagrangian_theta}
\mathcal{L}_{{\rm S}}(m,g)\to\mathcal{L}_{{\rm S}}(m,g,\theta)=\mathcal{L}_{{\rm S}}(m,g)+\frac{\theta}{2\pi}gF_{0,1}.
\eeq
 This vacuum $\theta$ angle is a c-number with a simple interpretation: it is proportional to an external background electric field, which is responsible for the  the origin of the vacua  degeneracy in the massless limit~\cite{schwinger_theta_vacuum,schwinger_theta_chiral}. In the massive regime $m\neq 0$, the Schwinger model can be used to understand charge shielding via the string tension between two separate probe charges (i.e. screening of the long-range Coulomb force between static charges)~\cite{massive_schwinger_shielding}, and string-breaking phenomena as the distance between the charges is increased beyond a certain value~\cite{string_breaking}. Moreover, the degeneracy with respect to the  $\theta$ angle is lifted, and one finds that for $\theta=\pi$ there is a continuous quantum phase transition between the confined phase with fermion trapping, and a distinct symmetry-broken phase with a so-called fermion condensate~\cite{massive_schwinger_theta}, which  is reminiscent of an exciton condensate from a condensed-matter perspective. 

\subsection{Lattice discretization  of the  Schwinger model}
There are  various  numerical methods to  unveil the above  non-perturbative phenomena,  and benchmark the analytical predictions. These methods typically rely on a discretization of the fermionic and gauge fields on a lattice~\cite{qcd_confinement}, and we shall focus on the Kogut-Susskind Hamiltonian approach~\cite{ham_lgt}. Here, only the spatial coordinates  are discretized into the sites of a chain $\Lambda_{\ell}=\{{\rm x}: {\rm x}/a\in\mathbb{Z}_{N_s}\}$, where $a$ is the lattice spacing, and $N_s$ is the number of lattice sites. By writing ${\rm x}=n a$ for $n\in\mathbb{Z}_{N_s}$, the matter sector of Dirac fermions can be represented by the so-called staggered fermions defined on the lattice sites $\Psi({\rm x}),\overline{\Psi}({\rm x})\to c_n^{\phantom{\dagger}}, c^{\dagger}_n$, such that $\{c_n^{\phantom{\dagger}},c_m^\dagger\}=\delta_{n,m}/a$, which have an alternating staggered mass $m_{\rm s}(-1)^n$ depending on the parity of the site. 

The gauge field sector, in the temporal gauge $A_0=0$, can be represented by rotor-angle operators (i.e. {\it compact QED}) living on the links of the lattice, and fulfilling $[L_n,\Theta_m]=-\ii\delta_{n,m}$. Here, the angle   operators are related to the gauge field $\Theta_n=agA_1({\rm x})$ at ${\rm x}=(n+\half) a$, while the rotors correspond to angular-momentum operators  related to the  electric field $L_n=E({\rm x})/g=F_{01}({\rm x})/g$. In this gauge, and using Schwinger's prescription for  gauge-invariant point-split operators $\bar{\Psi}({x}+\epsilon)\Psi({ x})\to\bar{\Psi}({ x}+\epsilon)\ee^{-\ii g\int_{{ x}}^{{ x}+\epsilon}{\rm d}x^{\mu}A_\mu(x)}\Psi({ x})$~\cite{schwinger_ps_gauge}, also known as the Peierls' substitution in condensed matter, the continuous-time Hamiltonian LGT for the standard massive Schwinger model becomes
\beq
\label{eq:standard_KS_schwinger}
H_{{\rm S}}\!=a\!\sum_{n=1}^{N_s} \!\left(\!\frac{-1}{2a}\!\left(\ii c_n^{{\dagger}}U_{n}^{\phantom{\dagger}}c_{n+1}^{\phantom{\dagger}}+{\rm H.c.}\!\right)\!+m_{\rm s}(-1)^nc_n^{{\dagger}}c_n^{\phantom{\dagger}}+\frac{g^2}{2}L_n^2\right)\!.
\eeq
Here, we have introduced the link operators $U_{n}=\ee^{\ii\Theta_n}$, which  act as unitary ladder operators $U_{n}\ket{\ell}=\ket{\ell+1}$ in the basis of electric-flux eigenstates $L_n\ket{\ell}=\ell\ket{\ell}$ for  $\ell\in\mathbb{Z}$. 

Finally, we note  that  the aforementioned vacuum angle~\eqref{eq:schwinger_lagrangian_theta} can be introduced in this Hamiltonian formulation through a background  electric field $E_{\rm ext}$  by substituting $L_n\to L_n+\theta/2\pi$, where $\theta=2\pi E_{\rm ext}/g$. With this notation, the lattice fields have the classical mass (energy) dimensions $d_{c}=1/2$ and $d_{L}=0$, while the mass and gauge coupling have $d_{m_s}=d_g=1$, and the lattice constant $d_a=-1$. In the continuum limit, one recovers Eq.~\eqref{eq:schwinger_lagrangian} with the staggered mass playing the role of the Dirac mass $
\mathcal{L}_{{\rm S}}(m_{\rm s},g,\theta)$.

\subsection{Topological Schwinger model, quantum $\hat{\theta}$ angle, and $\mathbb{Z}_N$ lattice gauge theories}
In our previous work~\cite{spt_schwinger}, we used an alternative discretization of the matter fields, which  hosts an SPT phase where the fermions interact via the gauge field. This alternative discretization follows from noting that the staggered mass in Eq.~\eqref{eq:standard_KS_schwinger} breaks translational invariance leading to a two-site unit cell. One may explore other discretizations with a two-site unit cell that also lead to massive fermions in the continuum limit. One possibility is to  dimerize the tunnelings, as occurs in  of the so-called Su-Schrieffer-Hegger model of polyacetylene in the static limit of a Peierls-dimerized chain~\cite{ssh_polyacetylene,polyacetylene_cont}. The gauge-invariant version of these dimerized tunnelings leads to the lattice version of the topological  Schwinger model
\beq
\label{eq:dimerised_KS_schwinger}
H_{t{\rm S}}\!=a\!\sum_{n=1}^{N_{\rm s}} \!\left(\!\frac{-1}{a} \left(\ii (1-\delta_n)c_n^{{\dagger}}U_{n}^{\phantom{\dagger}}c_{n+1}^{\phantom{\dagger}}+{\rm H.c.}\!\right)\!+\frac{g^2}{2}L_n^2\right)\!,
\eeq
where the dimerization vanishes for even sites $\delta_{2n}=0$, while it can be finite for odd sites $\delta_{2n-1}=\Delta$, and all the remaining operators and constants have been defined around Eq.~\eqref{eq:standard_KS_schwinger}. We note that the total number of sites $N_{s}$ should be even to respect inversion symmetry about the center of the chain.

As discussed in detail in Ref.~\cite{spt_schwinger}, the continuum limit of the matter sector~\cite{polyacetylene_cont} must be carefully reconsidered to correctly account for the possibility of having topological edge states and a non-zero topological invariant~\cite{zak_phase,berry_rmp}. Here, we simply summarise the main result: for $\Delta\in(0,2)$, one recovers Eq.~\eqref{eq:schwinger_lagrangian} in the continuum limit with a new term that depends on the fermion density of the topological edge states 
\beq
\label{eq:top_schwinger_lagrangian}
\mathcal{L}_{t{\rm S}}=\mathcal{L}_{t{\rm S}}(\Delta/a,g)-\sum_{\eta={\rm R,L}}\epsilon_{\eta}|\chi_\eta({\rm x})|^2\eta^\dagger\eta^{\phantom{\dagger}}.
\eeq
This continuum limit shows that the tunneling dimerization gives a non-zero mass $m=\Delta/a$ to the Dirac fermions, while the topological nature of the SPT phase is revealed by the topological edge states of energies $\epsilon_\eta$ and wavefunctions $\chi_\eta({\rm x})$, which are exponentially localised to the left- and right-most edges of the system. The bosonization of this continuum gauge theory must carefully account for the edge term, as it represents boundary charges that will change the boundary Gauss' law: the electric field can be discontinuous due to the charge localised at the left and right interfaces due to the existence of topological edge states. These considerations~\cite{spt_schwinger} lead to a quantum-mechanical edge contribution to the vacuum $\theta$ angle, which becomes an operator
\beq
\label{eq:theta_operator}
\hat{\theta}=\theta+\pi\left({\rm sign}(x){\rm L}^\dagger{\rm L}-{\rm sign}(x-N_{\rm s}a){\rm R}^\dagger{\rm R}\right).
\eeq
Such a contribution is physically reasonable. If only one of the edge states is populated by a fermion, there is an effective polarization of the material with a topological origin. Considering the theory of electromagnetic fields in dielectric media, such a polarization will modify the electric field inside the material. If the external electric field is such that $\theta=\pi$, we see that the population of a single edge state, and the associated polarization, does effectively cancel the electric field seen by the fermions. In situations where there is no  external electric field, the topological polarization of the media can self-generate a non-zero vacuum angle  $\theta=\pi$. 

\begin{figure}[t]
 \begin{centering}
  \includegraphics[width=0.95\columnwidth]{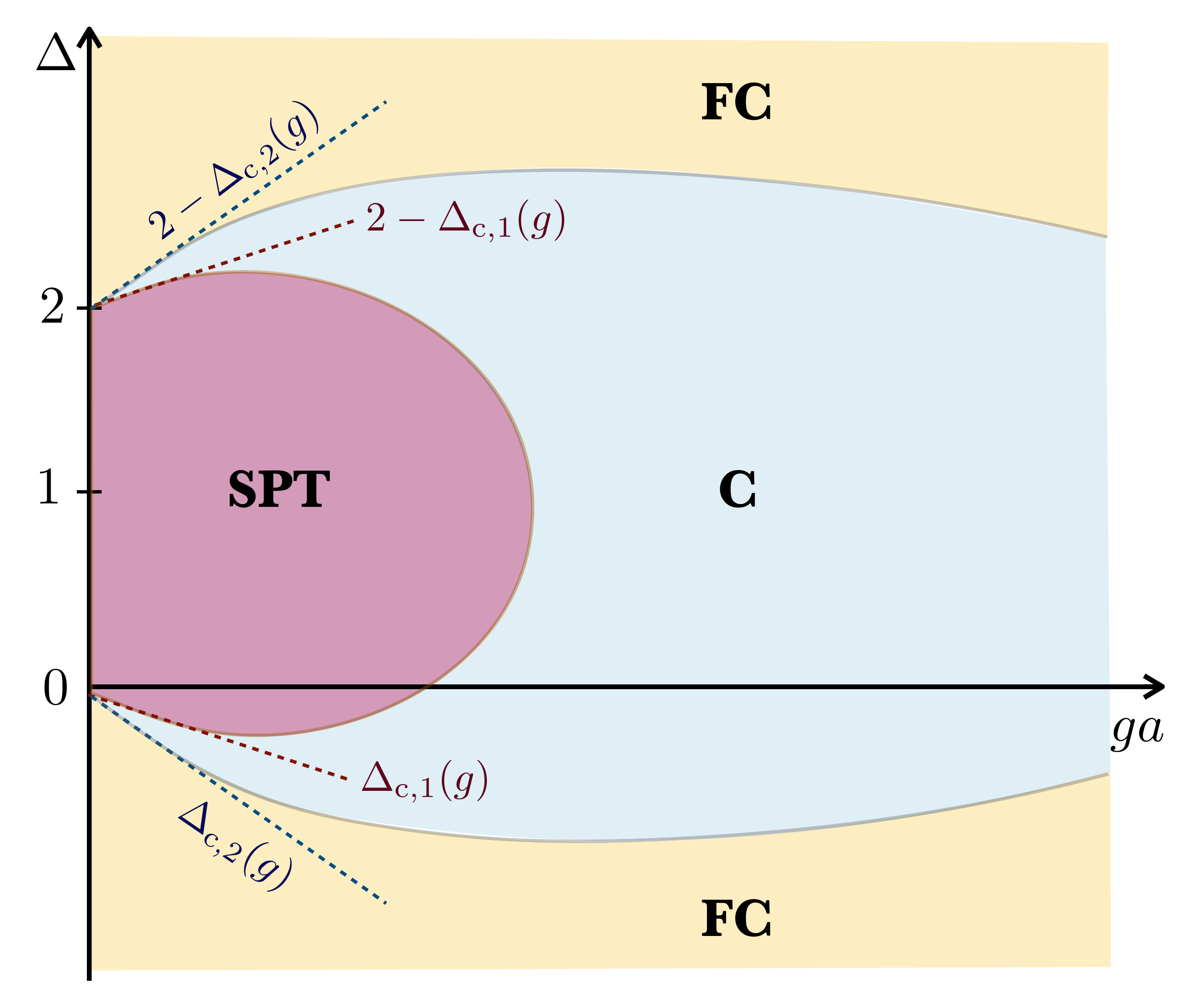}\\
  \caption{\label{Fig:qualitative_phase_diagram} {\bf Phase diagram of the  topological Schwinger model:} The bosonization with the  vacuum $\hat{\theta}$ operator~\eqref{eq:theta_operator}   allows us to predict three distinct phases: a symmetry-protected topological (SPT) phase, corresponding to a correlated topological insulator, separated from a confined phase (C) through a continuous quantum phase transition. This confined phase is itself separated from a symmetry-broken fermion condensate (FC) by another continuous  phase transition.}
\end{centering}
\end{figure}

This behaviour can certainly modify the  phase diagram  of the topological Schwinger model in comparison to the standard one. In particular, it leads to the phase diagram sketched in Fig.~\ref{Fig:qualitative_phase_diagram}. As outlined above Eq.~\eqref{eq:top_schwinger_lagrangian}, the vertical axis at zero gauge coupling separates an SPT groundstate for dimerizations $\Delta\in(0,2)$ from a trivial band insulator for $\Delta<0$ or $\Delta>2$. As the gauge coupling $g$ is switched on, the fermions in the SPT phase start interacting via the gauge bosons, and acquire correlations. As a consequence of such interactions, the bare mass $m_{\rm s}=\Delta/a$ in Eq~\eqref{eq:top_schwinger_lagrangian} gets renormalised, such that one finds  quantum phase transitions separating the SPT phase from a topologically-trivial phase. As can be observed in Fig.~\ref{Fig:qualitative_phase_diagram},  this phase corresponds to the  confined phase (C) displaying massive neutral quasiparticles composed of bound fermion-antifermion pairs, which was mentioned in the introduction. At this point, since there are no longer edge states, the vacum angle becomes $\theta=\pi$, where another quantum phase transition towards a parity-breaking fermion condensate (FC) can take place~\cite{massive_schwinger_theta}. This is precisely the second critical line displayed in Fig.~\ref{Fig:qualitative_phase_diagram}. We note that the bosonization allows for the analytical prediction of these two critical lines $\Delta=  \Delta_{\rm c,1}(g)$, and  $\Delta=  \Delta_{\rm c,2}(g)$, at weak couplings 
\beq
 \label{eq:spt_c_line_1}
 \Delta_{\rm c,1}(g)=-\frac{ga}{2\sqrt{\pi}}\ee^{-\gamma},
 \eeq
 where $\gamma\approx0.5774$ is Euler's constant, and 
 \beq
 \label{eq:fc_c_line_2}
\Delta_{\rm c,2}(g)=-\frac{1}{3}ga,
\eeq
together with their symmetric counterparts $\Delta= 2- \Delta_{\rm c,1}(g)$, and $\Delta= 2- \Delta_{\rm c,2}(g)$.

\begin{figure*}
 \begin{centering}
  \includegraphics[width=1.99\columnwidth]{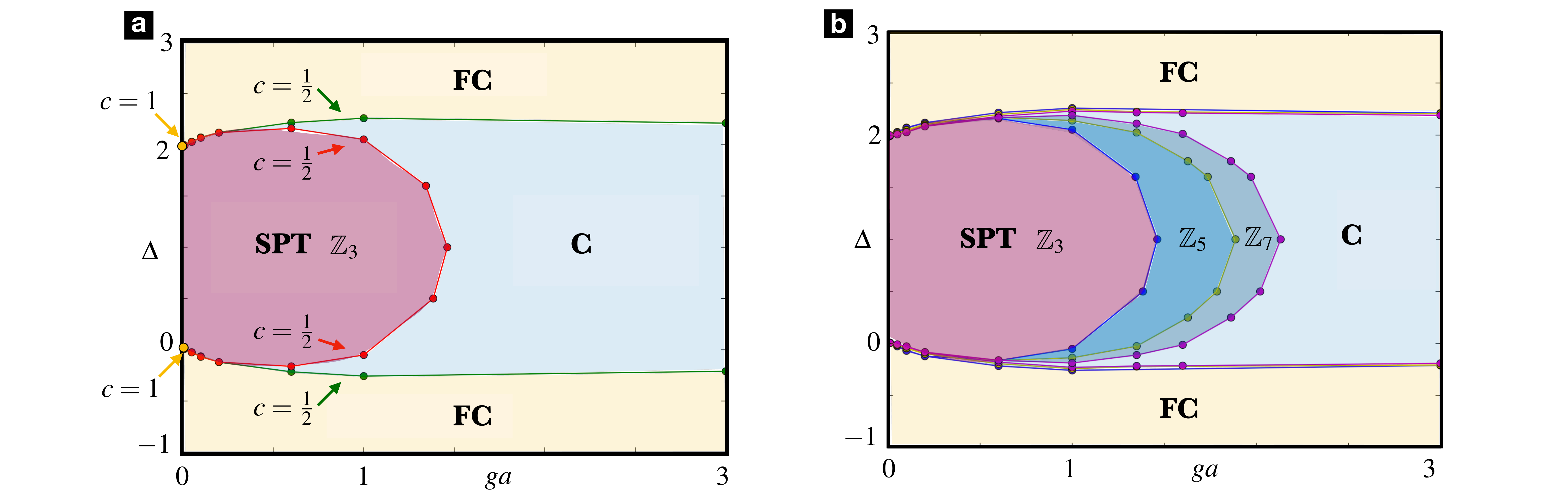}\\
  \caption{\label{Fig:phase_diagram_num} {\bf DMRG phase diagram for  $\mathbb{Z}_N$ topological Schwinger models:} {\bf (a)}  $\mathbb{Z}_{3}$ model, including the central charges of the critical lines. {\bf (b)} $\mathbb{Z}_{5}$ and $\mathbb{Z}_{7}$ models, showing that the extension of the SPT phase  grows as $N$ is increased.}
\end{centering} 
\end{figure*}
Let us now introduce the discrete $\mathbb{Z}_N$ gauge group version of the topological Schwinger model~\cite{spt_schwinger}, which shall be used to benchmark the bosonization predictions of Fig.~\ref{Fig:qualitative_phase_diagram}, and Eqs.~\eqref{eq:spt_c_line_1}-\eqref{eq:fc_c_line_2}. The Hamiltonian of $\mathbb{Z}_N$ LGT with topological matter reads
\beq
\label{eq:dimerised_KS_schwinger_Z_N}
H^{\mathbb{Z}_N}_{t{\rm S}}\!=a\!\sum_{n=1}^{N_s} \!\left(\!\frac{-1}{a} \left(\ii (1-\delta_n)c_n^{{\dagger}}\tilde{U}_{n}^{\phantom{\dagger}}c_{n+1}^{\phantom{\dagger}}+{\rm H.c.}\!\right)\!+\frac{g^2}{2}\tilde{L}_n^2\right)\!,
\eeq
where the main difference with the Kogut-Susskind approach~\eqref{eq:dimerised_KS_schwinger} is that the gauge fields are defined through the pair of link operators $\tilde{U}_n,\tilde{V}_n$ that obey the $\mathbb{Z}_N$  algebra, fulfilling $\tilde{U}_{n}^N=\tilde{V}_{n}^N=\mathbb{I}$, and $\tilde{V}^\dagger_{n}\tilde{U}_n\tilde{V}_{n}=\ee^{\ii2\pi/N}\tilde{U}_n$ ~\cite{alternative_z_n_gauge, zn_presentation, zn_study}.

By using the electric-flux eigenbasis $\tilde{V}_n\ket{v}=v\ket{v}$ with $v\in\mathbb{Z}_N$  on each link, one can understand that the link operator $\tilde{U}_{n}$ acts as ladder operator that raises the electric flux by one quantum $\tilde{U}_n\ket{v}=\ket{v+1}$ in a cyclic way, i.e.  $\tilde{U}_n\ket{N}=\ket{1}$. We note that these link operators can be defined in terms of the vector potential and the electric field  $\tilde{U}_n={\rm exp}\{\ii a g A_n\}$,   $\tilde{V}_n={\rm exp}\{\ii \frac{2\pi}{ N}\frac{E_n}{g}\}$. Accordingly, the $\mathbb{Z}_N$ algebra $[\tilde{U}_n,\tilde{V}_n]=\ee^{\ii 2\pi/N}$ can be  satisfied by imposing the usual  canonical commutation relations on the gauge fields  $[E_n,A_m]=\ii\delta_{n,m}/a$, which have the correct continuum limit $[E(x),A(y)]=\ii\delta(x-y)$.  Note also that the gauge-group condition $\tilde{U}_n^N=\tilde{V}_n^N=\mathbb{I}$ requires that the electric-flux eigenvalues of $\tilde{L}_n=E_n/g$ should span $\sigma(\tilde{L}_n)=\{-\half(N-1),\cdots,\half(N-1)\}$. This yields $\sigma(\tilde{L}_n)\to \mathbb{Z}$ in the large-$N$ limit, which corresponds to the spectrum of the rotor operator $L_n$ of the Kogut-Susskind approach ~\cite{ham_lgt}. In the same manner, the eigenvalues of the vector potential should lie in $\sigma(ag A_n)=\{-\pi(N-1)/N,\cdots,\pi(N-1)/{N}\}\to[-\pi,\pi]$, corresponding to the basis of the angle operator  $\Theta_n$ in the Kogut-Susskind approach,  and leading to compact QED$_2$. Therefore, one expects that the properties of compact QED$_2$ coupled to a topological matter sector can be reproduced by taking the large-$N$ limit of these $\mathbb{Z}_N$ LGTs.  

Let us now address an important point of the lattice formulation. In the continuum, Gauss' law and its interplay with bound topological charge at the edges is of paramount importance to determine the phase diagram. In the lattice, in order to take into account  Gauss's law,  we introduce the operator 
\beq
\label{eq:magnifico_gauss_law}
G_{n}=c_{n}^{\dagger}c_{n} + \frac{1}{2a}[(-1)^{n}-1]-\frac{1}{a}(\tilde{L}_{n}-\tilde{L}_{n-1}).
\eeq
Accordingly,  $\left |  \psi \right > $ is a physical state if it satisfies the condition 
\beq
\label{eq:gauss_law}
G_{n}\left |  \psi \right > = 0 \hspace{2ex} \forall n\in\{1,2,\cdots,N_{\rm s}\}.
\eeq
This is a very important constraint that allows us to construct the physical Hilbert space of the $\mathbb{Z}_{N}$ model, as we will see in the following section.

\section{\bf Density-matrix-renormalization-group  simulations of the topological Schwinger model}
\label{sec:dmrg}

\subsection{$\mathbb{Z}_3$ topological Schwinger model}

In this section, we start by reviewing the DMRG phase diagram for the simplest non-trivial case presented in~\cite{spt_schwinger}, the $\mathbb{Z}_{3}$ model~\eqref{eq:dimerised_KS_schwinger_Z_N} with three electric-flux levels on each link. This will serve us to set the common ground for the large-$N$ studies presented in the following sections. Before turning to larger values of $N$, we  discuss new results concerning the entanglement spectroscopy~\cite{entanglemnent_review} for the critical lines of the $\mathbb{Z}_3$ LGT.

The different phases depicted in Fig.~\ref{Fig:qualitative_phase_diagram} can be characterised by two different observables: the {\it electric-field order parameter}, which can be defined as
\beq
\label{eq:magnifico_electric_parameter}
\Sigma = \frac{1}{N_s}\sum_{n=1}^{N_{s}} \bra{\rm gs} E_{n}  \ket{\rm gs}=\frac{g}{N_s}\sum_{n=1}^{N_{s}} \bra{\rm gs} \tilde{L}_{n}  \ket{\rm gs}
\eeq
and the {\it topological correlator}, namely
\beq
\label{eq:top_correlator}
O_{-} =\frac{2}{N_{\rm s}}\sum_{n=1}^{N_{\rm s}/2} O^{(2n-1)}_{-},
\eeq
where we sum over all odd sites of the chain the observables
\beq
\label{eq:magnifico_topological_order_quantity}
O^{(n)}_{-} = \frac{3}{2}\bra{\rm gs} c_{n}^{\dagger}c_{n+1}^{\phantom{\dagger}} +  c_{n+1}^{\dagger}c_{n}^{\phantom{\dagger}}  \ket{\rm gs} + \rho_{n, n+1} -  \frac{1}{2}(\rho_{n}+\rho_{n+1}).
\eeq
Here,  $\rho_{n}=\bra{\rm gs}c^\dagger_nc^{\phantom{\dagger}}_n\ket{\rm gs}$  are the densities of fermions, while  $ \rho_{n, n+1}= a\bra{\rm gs} c^\dagger_nc^{\phantom{\dagger}}_nc^\dagger_{n+1}c^{\phantom{\dagger}}_{n+1}\ket{\rm gs}$ represent the density-density correlations, and the site index $n$ must be odd. Such a topological correlator  was introduced for the dimerized free-fermion model~\cite{ssh_order_parameters}, and used in Ref.~\cite{spt_schwinger} to map the phase diagram of the full $\mathbb{Z}_3$ topological Schwinger model. 

\begin{figure}[t]
 \begin{centering}
  \includegraphics[width=0.8\columnwidth]{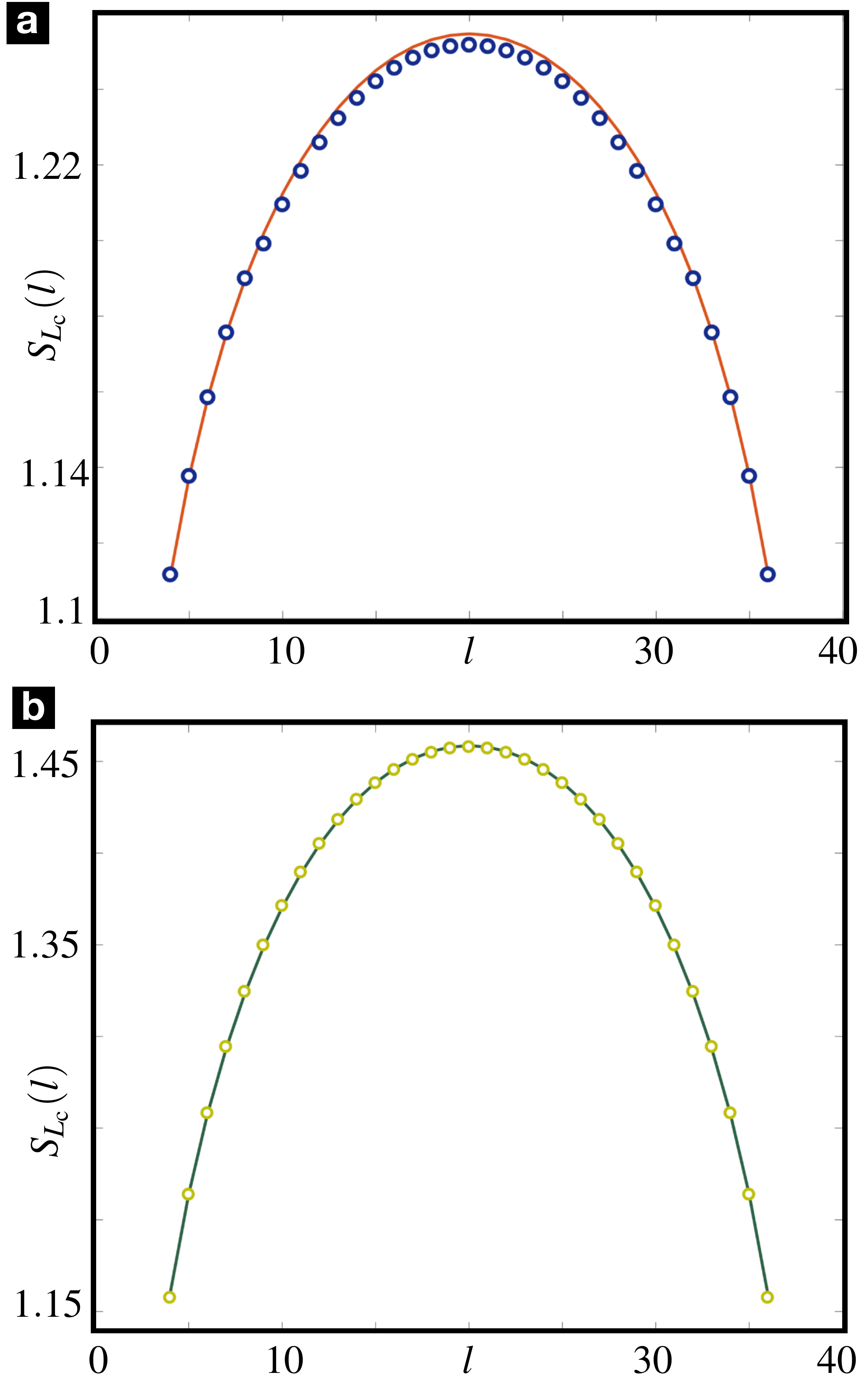}\\
  \caption{\label{Fig:ent_entropy} {\bf Block entanglement entropy of the $\mathbb{Z}_3$  model:}     {\bf (a)} Scaling of the entanglement entropy of a subsystem of size $l$ on the critical point $\Delta=-0.162$ and $ga=0.6$. Through a logarithmic fit~\eqref{eq:magnifico_scaling_entropy}, it is possible to extract the central charge $c=0.506$.   {\bf (b)} Same as {\bf (a)}, but  for $g=\Delta=0$.}
\end{centering}
\end{figure}

By studying the behavior of these two quantities and performing a careful finite-size scaling analysis, we find the phase diagram of the the $\mathbb{Z}_{3}$ topological Schwinger model shown in Fig.~\ref{Fig:phase_diagram_num} {\bf (a)}. The three different phases can be characterised as follows: i) the {\it confined phase} (C) with $\Sigma=0$ and $O_{-} < 0$; ii) the {\it fermion condensate} (FC) with $\Sigma > 0$ and $O_{-} < 0 $; iii) the symmetry-protected topological (SPT) phase with $\Sigma < 0 $ and $O_{-} > 0$. We recall that  the topological nature of SPT phase was revealed in~\cite{spt_schwinger} by showing the  exact degeneracy in the { entanglement spectrum}~\cite{ent_spectrum}, and  the existence of many-body edge states. In addition, from the scaling analysis, we estimated the values of the critical exponents $\beta$ and $\nu$, related to the two phase transitions FC-C and SPT-C. In both cases, we found $\beta=1/8$ and $\nu=1$, which correspond to the 2D Ising universality class. 

We want now to confirm this finding by performing another independent analysis based on the scaling of the { entanglement entropy}.  The entanglement entropy $S(\tilde{\rho}_{A})=-{\rm Tr} \left [ \tilde{\rho}_{A}\log_2\left (\tilde{\rho}_{A}\right )\right ]$ is defined for the reduced density matrix of a partition $A$ of our system $\tilde{\rho}_{A}={\rm Tr}_{B}{\ket{\rm gs}\bra{\rm gs}}$, where $B$ is the complement of $A$. According to conformal field theory (CFT)~\cite{entan_entropy,entanglement_entropy_calabrese_cardy}, considering a subsystem  A of size $l$ within the chain that has $L_{\rm c}=N_{\rm s}/2$ couples of sites, we expect to observe  a logarithmic scaling of the block entanglement entropy if the system is at a quantum critical point
\beq
\label{eq:magnifico_scaling_entropy}
S_{L_{\rm c}}(l)= \frac{c}{6}\log_2\left [\frac{2L_{\rm c}}{\pi}\sin \left ( \frac{\pi l}{L_{\rm c}} \right ) \right ] + s_{0},
\eeq
where $s_{0}$ is a non-universal constant, and $c$ is the central charge of the corresponding CFT that governs the critical behavior. As shown in Fig.~\ref{Fig:ent_entropy}{\bf (a)}, we observe such a logarithmic scaling for   the critical point  $ga=0.6$ and $\Delta=-0.162$, from which it is possible to extract the central charge through a logarithmic fit. We obtain the value $c=0.506$, in  agreement with the central charge of 2D Ising universality class $c =\half$. Interestingly, by switching off the gauge coupling $g=0$, we obtain the entanglement entropy of Fig.~\ref{Fig:ent_entropy}{\bf (b)} at the critical point $\Delta=0$, which yields $c=1.059$ through the logarithmic fit. This result is in agreement with the expectation for the critical point of the dimerised free-fermion model, which can be described by the CFT of a massless Dirac fermion with $c=1$. Accordingly, the $c=1$ CFT  of the non-interacting  model splits into a couple of  CFTs with $c=\half$, each of which controls the criticality of the SPT-C and C-FC  quantum phase transitions, as depicted in Fig.~\ref{Fig:phase_diagram_num} {\bf (a)}. We note that a similar splitting of a $c=1$ massless Dirac fermion into a couple of $c=1/2$ massless Majoranas has been recently observed in other models of correlated SPT phases with instantaneous   Hubbard-type interactions instead of the gauge-mediated ones~\cite{creutz_hubbard_ladder,gross_neveu_wilson, wilson_hubbard_matter}.

\begin{figure}[t]
 \begin{centering}
  \includegraphics[width=0.98\columnwidth]{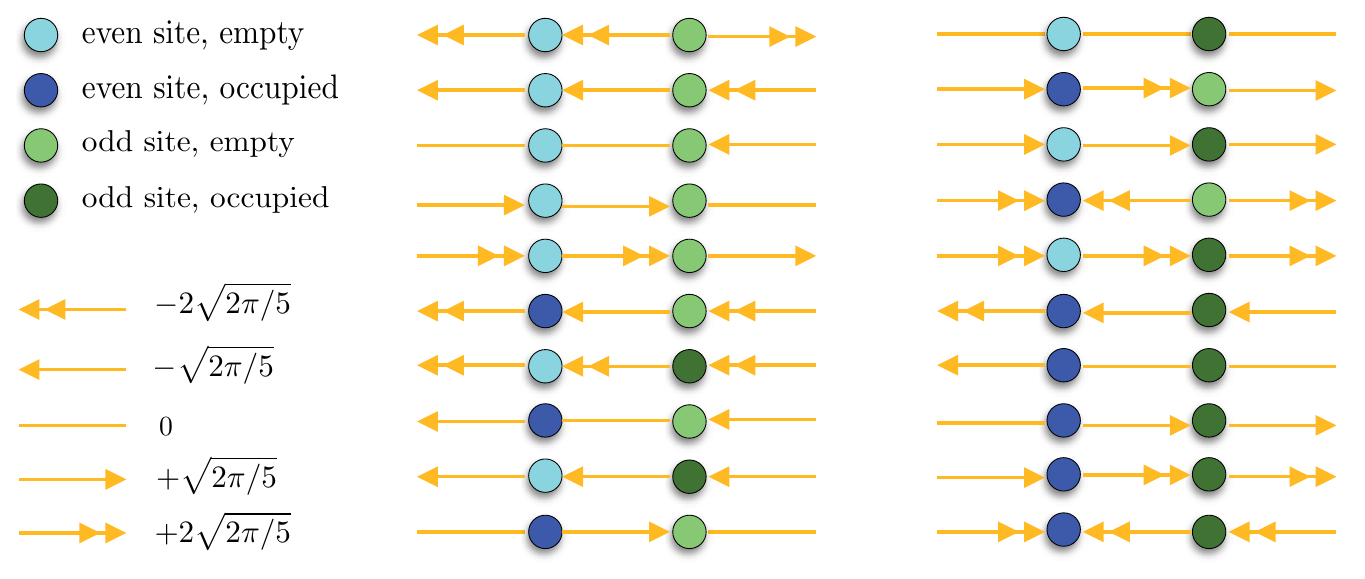}\\
  \caption{\label{Fig:basis_z5} {\bf Basis of gauge-invariant Hilbert space for the $\mathbb{Z}_5$ model:}  taking into account Gauss's law, we implement all possibile configurations of matter/antimatter fields on even/odd sites.}
\end{centering}
\end{figure}

\begin{figure*}[t]
 \begin{centering}
  \includegraphics[width=1.6\columnwidth]{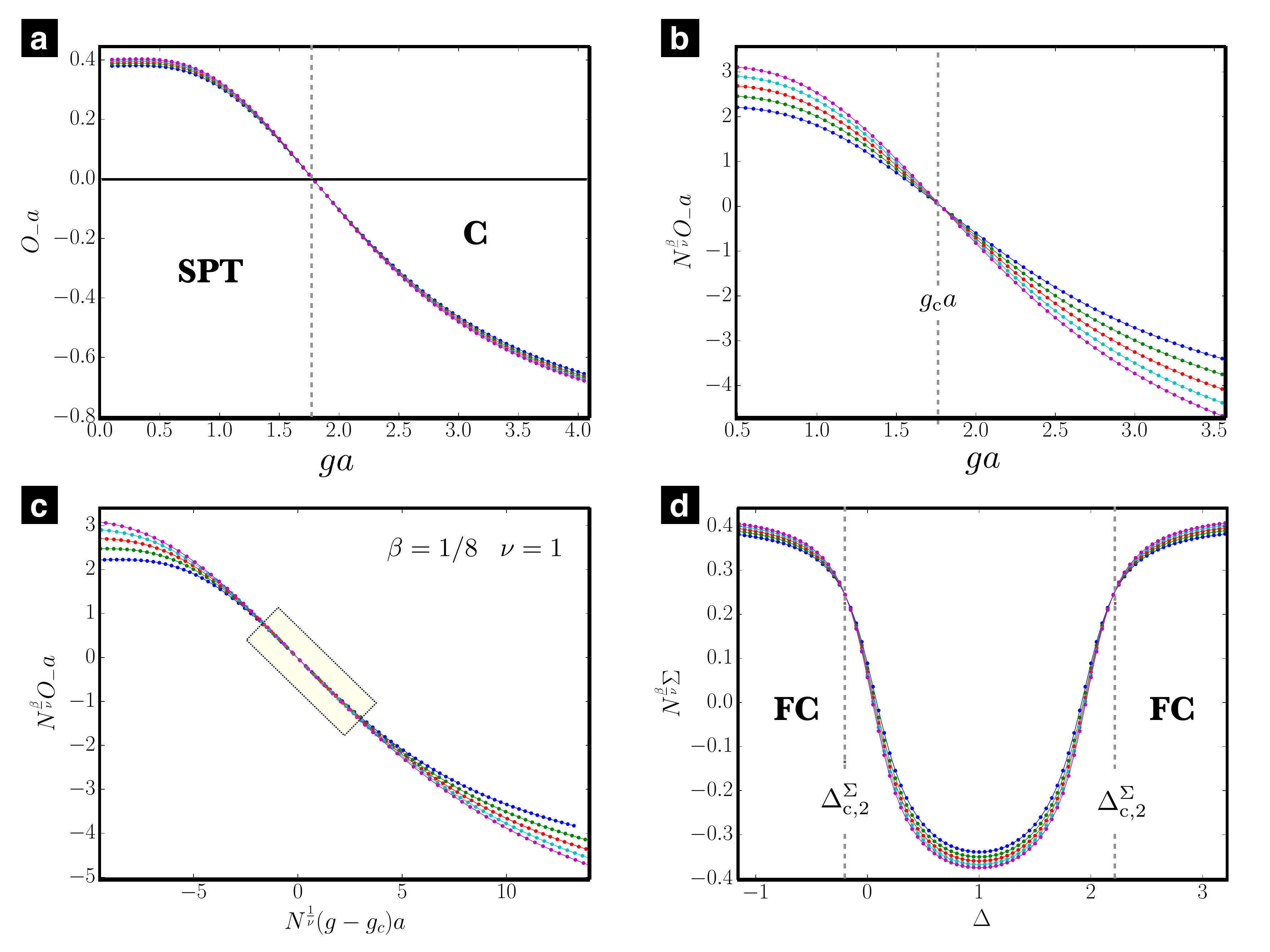}\\
  \caption{\label{Fig:all_plots_z5}{\bf Observables and scaling relations for the $\mathbb{Z}_5$ topological Schwinger model:} {\bf (a)} Behavior of the topological correlator for $\Delta=0.5$ as a function of the gauge coupling constant g with $N_{s}=\left[40,50,60,70,80\right]$ sites (from blue ($40$) to purple ($80$)). {\bf (b)} Scaling quantity $N^\frac{\beta}{\nu}O_{-}a$ of the topological correlator calculated for $\Delta=0.5$ as a function of the gauge coupling for various system sizes $N_{s}=\left[24, 28, 32, 36, 40\right]$ (from blue ($24$) to purple ($40$)). The crossing point of all curves allows us to determine the critical point separating the SPT and C phases $g_{c}a\approx 1.786$. {\bf (c)} Universal scaling of the topological correlator with the critical exponents of the 2D Ising universality class $\beta =1/8$ and $\nu=1$: all the numerical curves for different system sizes $N_{s}$ collapse onto the same universal function $\lambda(x)$ in the shaded region. {\bf (d)} Scaling quantity $N^\frac{\beta}{\nu}\Sigma$ of the electric field order parameter calculated for $ga=0.6$ as a function of the dimerization for various system sizes $N_{s}=\left[24, 28, 32, 36, 40\right]$ (from blue ($24$) to purple ($40$)). The crossing point of all curves allows us to determine the critical point separating the FC and C phases $\Delta^{\Sigma}_{c}\approx -0.193$ and $\Delta^{\Sigma}_{c} \approx 2.105 $ in accordance with the symmetry around $\Delta=1$.}
\end{centering}
\end{figure*}

 \subsection{ $\mathbb{Z}_5$ and $\mathbb{Z}_7$ topological Schwinger model}
 
In this section, we  consider the following question: what is the fate of the SPT phase as we increase the dimension of the gauge group symmetry $\mathbb{Z}_N$? To address this question, we  perform DMRG simulations of our topological LGT~\eqref{eq:dimerised_KS_schwinger_Z_N} for larger values of $N$. We start by considering the $\mathbb{Z}_5$ case with five possible electric field levels on each link. In order to impose Gauss' law~\eqref{eq:gauss_law} with the corresponding operator~\eqref{eq:magnifico_gauss_law}, one can directly build the gauge-invariant Hilbert space considering certain local configurations. For the $\mathbb{Z}_5$ gauge group, the basis of such local gauge-invariant configurations can be built by considering the two-site unit cell, with the corresponding gauge fields,  as shown in Fig.~\ref{Fig:basis_z5}. Through this local basis, we directly implement the gauge-invariant subspace in our DMRG code, reducing the complexity of the problem. We determine the ground-state of the system $\left |{\rm gs} \right >$ with open boundary conditions, working with up to $N_{s} = 80$ sites and keeping 1200 DMRG-states at most. These values are large enough to ensure stability of our findings and small truncation errors.

First of all, in order to prove the presence of the SPT phase also in the $\mathbb{Z}_5$ model, we analyze the behavior of the topological correlator $O_{-}$~\eqref{eq:top_correlator} for $\Delta=0.5$ as a function of the gauge coupling $g$. As shown in Fig.~\ref{Fig:all_plots_z5}{\bf(a)}, we obtain a clear sign reversal of the correlator, a behavior that is qualitatively analogous to the transition from the topological phase ($O_{-} > 0$) to the trivial one ($O_{-} < 0$)  observed in the  $\mathbb{Z}_3$ case~\cite{spt_schwinger}. In order to extract the critical point, we can perform a finite-size scaling of the SPT order parameter $O_{-}$. In the  $\mathbb{Z}_3$ model, the quantum phase transition SPT-C has critical exponents $\beta=1/8$ and $\nu=1$. Therefore, we can start to explore the critical behavior by assuming these values in the usual scaling relation
\beq
\label{eq:magnifico_topological_scaling_relation}
N_{\rm s}^{\frac{\beta}{\nu}}O_{-} = \lambda \left ( N_{\rm s}^{\frac{1}{\nu}} (g-g_{\rm c}) \right ).
\eeq
 in which $\lambda(x)$ is a universal function. By fixing  $\Delta=0.5$ and plotting the quantity $N_{\rm s}^{\beta/\nu}O_{-}$ as a function of $g$ for different values of $N_{s}$, we obtain the behavior of Fig.~\ref{Fig:all_plots_z5}{\bf (b)}. From the scaling relation \eqref{eq:magnifico_topological_scaling_relation}, it results that, for $g=g_{\rm c}$, the value  $\lambda(0)$ becomes independent of the system size. Therefore, one should observe a crossing of the curves for different lengths precisely at the critical point. This is exactly the behavior displayed in Fig.~\ref{Fig:all_plots_z5}{\bf (b)}, wich has a clear crossing that yields a value of the critical point of $g_{\rm c} \approx 1.786/a$. We now have to look at the numerical curves given by $N^{\beta/ \nu}_{s}O_{-}$ versus $N^{1/\nu}_{s}(g-g_{c})$, for different $N_{s}$, which should all collapse onto the same universal function $\lambda(x)$ near the origin in case the critical exponents coincide with those of the 2D Ising model. This represents an important check for the values of the critical exponents, as clearly visible in Fig.~\ref{Fig:all_plots_z5}{\bf (c)}.
 
Following this scheme, we can iterate this procedure i) for different values of $\Delta$ (horizontal lines in the plane $ga-\Delta$); ii) fixing a particular value of $g$ and varying the dimerization parameter $\Delta$ (vertical lines in the plane $ga-\Delta$). In this way we determine the critical values $\Delta^{O_{-}}_{\rm c,1}$ for  the transitions SPT-C shown in Tab.~\ref{tab:table_of_critical_values_vertical_z5}, corresponding to the lower and upper parts of the critical line depicted in Fig.~\ref{Fig:qualitative_phase_diagram}. As in the $\mathbb{Z}_3$ model, when the gauge coupling $g$ is sufficiently large, this transition is absent. We also observe the symmetry around  $\Delta=1$.
 
We can now investigate the behavior of the FC-C transition by using the electric field order parameter $\Sigma$~\eqref{eq:magnifico_electric_parameter}. Assuming the same critical exponents $\beta=1/8$, $\nu=1$,  we perform a finite-size scaling analysis with
\beq
\label{eq:magnifico_electric_scaling_relation}
N_{\rm s}^{\frac{\beta}{\nu}}\Sigma = \lambda \left ( N_{\rm s}^{\frac{1}{\nu}} (\Delta-\Delta_{\rm c}) \right ),
\eeq
 Accordingly,  for $g=0.6$, we obtain the plot in Fig.~\ref{Fig:all_plots_z5}{\bf(d)}, which allows us to detect two critical points $\Delta_{\rm c,1}^{\Sigma} \approx -0.193$ and $\Delta_{\rm c},1^{\Sigma} \approx 2.105$, again symmetrical with respect to $\Delta=1$. 

Also in this case, we can iterate this procedure to determine the critical points related to the transition FC-C for different values of $g$. The resulting values $\Delta^{\Sigma}_{\rm c,2}$ are shown in Table~\ref{tab:table_of_critical_values_vertical_z5}, and correspond to the to the lower and upper parts of the critical line depicted in Fig.~\ref{Fig:qualitative_phase_diagram}.
\begin{table}
\centering{}%
\begin{tabular}{|c|c|c|c|c|}
\hline 
$ga$ & $\Delta_{\rm c,2}^{\Sigma}$ & $\Delta_{\rm c,2}^{\Sigma}$ & $\Delta_{\rm c,1}^{O_{-}}$ & $\Delta_{\rm c,1}^{O_{-}}$\tabularnewline
\hline 
\hline 
$0.01$ & $0.005$ & $1.998$ & $0.007$ & $1.996$\tabularnewline
\hline 
$0.05$ & $-0.016$ & $2.015$ & $-0.015$ & $2.018$\tabularnewline
\hline 
$0.10$ & $-0.045$ & $2.050$ & $-0.042$ & $2.045$\tabularnewline
\hline 
$0.20$ & $-0.098$ & $2.105$ & $-0.096$ & $2.097$\tabularnewline
\hline 
$0.60$ & $-0.193$ & $2.198$ & $-0.162$ & $2.170$\tabularnewline
\hline 
$1.00$ & $-0.240$ & $2.246$ & $-0.136$ & $2.140$\tabularnewline
\hline 
$1.35$ & $-0.220$ & $2.225$ & $-0.025$ & $2.025$\tabularnewline
\hline 
$3.00$ & $-0.206$ & $2.207$ & $//$ & $//$\tabularnewline
\hline 
\end{tabular}\caption{\label{tab:table_of_critical_values_vertical_z5}{\bf  $\mathbb{Z}_{5}$ topological Schwinger model:} critical values of $\Delta$ (related to the two transitions FC-C and SPT-C) obtained for different values of $g$. The numerical error is equal to $10^{-3}$.}
\end{table}

We repeat all the numerical analysis also for the $\mathbb{Z}_{7}$ model (i.e. seven possible electric field levels on each link), observing the same behavior of the parameters $O_{-}$ and $\Sigma$. The critical values of both transitions SPT-C and FC-C are reported in Tab.~\ref{tab:table_of_critical_values_vertical_z7}. Putting together all these numerical results, we obtain the phase diagrams of the $\mathbb{Z}_{5}$ and $\mathbb{Z}_{7}$ topological Schwinger model shown in Fig.~\ref{Fig:phase_diagram_num} with a comparison with the $\mathbb{Z}_{3}$ case. Here, one can observe that the extension of the SPT phase grows with $N$ while, simultaneously, the spacing between the critical lines becomes smaller. We can thus answer the question raised at the beginning of this section: as $N$ increases, the correlated SPT phase remains stable, and exits for a wider region of parameter space by displacing the confined phase towards larger values of the gauge coupling $g$. In the following section, we will address the large-$N$ limit, benchmarking the results for the $U(1)$ LGT by the bosonization predictions.

\begin{table}
\centering{}%
\begin{tabular}{|c|c|c|c|c|}
\hline 
$ga$ & $\Delta_{\rm c,2}^{\Sigma}$ & $\Delta_{\rm c,2}^{\Sigma}$ & $\Delta_{\rm c,1}^{O_{-}}$ & $\Delta_{\rm c,1}^{O_{-}}$\tabularnewline
\hline 
\hline 
$0.01$ & $0.004$ & $1.998$ & $0.006$ & $1.997$\tabularnewline
\hline 
$0.05$ & $-0.010$ & $2.009$ & $-0.008$ & $2.007$\tabularnewline
\hline 
$0.10$ & $-0.030$ & $2.032$ & $-0.027$ & $2.028$\tabularnewline
\hline 
$0.20$ & $-0.085$ & $2.089$ & $-0.082$ & $2.084$\tabularnewline
\hline 
$0.60$ & $-0.178$ & $2.181$ & $-0.160$ & $2.168$\tabularnewline
\hline 
$1.00$ & $-0.228$ & $2.231$ & $-0.186$ & $2.190$\tabularnewline
\hline 
$1.35$ & $-0.215$ & $2.218$ & $-0.110$ & $2.110$\tabularnewline
\hline 
$1.60$ & $-0.212$ & $2.212$ & $-0.012$ & $2.012$\tabularnewline
\hline 
$3.00$ & $-0.190$ & $2.190$ & $//$ & $//$\tabularnewline
\hline 
\end{tabular}\caption{\label{tab:table_of_critical_values_vertical_z7}{\bf  $\mathbb{Z}_{7}$ topological Schwinger model:} critical values of $\Delta$ (related to the two transitions FC-C and SPT-C) obtained for different values of $g$. The numerical error is equal to $10^{-3}$.}
\end{table}

\begin{table}
\centering{}%
\begin{tabular}{|c|c|c|c|}
\hline 
$ $ & $\mathbb{Z}_{3}$ & $\mathbb{Z}_{5}$ & $\mathbb{Z}_{7}$\tabularnewline
\hline 
\hline 
$\Delta_{c}^{\Sigma}$ & $-0.67475248$ & $-0.54405941$ & $-0.48118812$\tabularnewline
\hline 
$\Delta_{c}^{O_{-}}$ & $-0.66831683$ & $-0.54158416 $ & $ -0.46237624$\tabularnewline
\hline 
\end{tabular}\caption{\label{tab:different_slopes}Different slopes for small $g$ of the two critical lines.}
\end{table}

\subsection {Large-$N$  phase diagram and topological QED$_2$}

From the previous section, the extent of the SPT phase clearly grows as the dimension of the gauge group is increased with  $N$.
In order to access the $U(1)$ limit of the $\mathbb{Z}_{N}$ topological Schwinger model~\eqref{eq:dimerised_KS_schwinger_Z_N}, and determine if the SPT phase is finite or if it completely removes the confined phase, we can study the scaling with $N$ of the outer critical point $g_{\rm c}$ at the tip of the SPT lobe  by fixing $\Delta=1$. By fitting the critical points with an exponential function $
g_{\rm c} (N)a= A e^{-B/N} + C
$, we 
obtain the fitting parameters $A \approx 2.323$,   $B \approx 3.177$,   $C \approx 0.656$. In this way, we can extract a finite critical value in the $N \rightarrow \infty$ limit
$
g_{\rm c} (\infty)a = A + C \approx 2.979,
$
which shows that the SPT phase survives to considerably strong gauge interactions $ga\approx 3$, but eventually gives way to the confined phase. 

Similarly, we can explore the vertical extent of the SPT phase by fitting  the  critical points $\Delta_{\rm c,1}^{O_{-}}$ for $ga=0.2$  as a function of $N$. Due to the symmetry about $\Delta=1$, it suffices to consider the lower of the two symmetrical critical lines. In this case, we obtain the  extrapolation $\Delta_{\rm c,1}(\infty) \approx   -0.033$. In light of this result, we can conclude that the SPT phase has a finite region of stability  in the presence of gauge couplings $g > 0$, which is in accordance to the analytical results that we obtained for the $U(1)$ topological Schwinger model (see Fig.~\ref{Fig:qualitative_phase_diagram}). In this sense, our numerical results manifest the expectation that the $\mathbb{Z}_N$ theory  yields the $U(1)$ LGT in  the limit $N \rightarrow \infty$, in which the electric field can assume any continuous value ($\mathbb{Z}_{N} \rightarrow U(1)$). 

In order to take one step further in this comparison, and provide a quantitative benchmark of the analytic $U(1)$ results, we now analyze the slope of the the two critical lines (SPT-C and FC-C) for small $g$. This will allow us to test the bosonization predictions in Eqs.~\eqref{eq:spt_c_line_1}-\eqref{eq:fc_c_line_2}. For each $\mathbb{Z}_{N}$ model, we have calculated the critical points $\Delta_{\rm c}^{\Sigma}$ and $\Delta_{\rm c}^{O_{-}}$ for $ga=0.01,0.05,0.10,0.20$, and performed  a linear fit to extract the different slopes $m(N)$. We obtain the values in Table~\ref{tab:different_slopes}, which can be fitted to a function of $N$ with an exponential behavior of the form $m(N)=\zeta e^{-(\tau/N)}+\kappa$. This allows us to obtain an extrapolation of the slopes of the two critical lines (SPT-C and FC-C) in the limit $N \rightarrow \infty$
\beq
\label{eq:magnifico_slopes_extrapolation} \begin{split}
m_{1}(\infty) = \zeta_{1} + \kappa_{1} =  -0.1625 \\
m_{2}(\infty) = \zeta_{2} + \kappa_{2} = -0.3034.
\end{split}
\eeq
These values are in remarkable agreement with  the expected ones derived for the $U(1)$ limit in Eqs.~\eqref{eq:spt_c_line_1}-\eqref{eq:fc_c_line_2} (respectively $-e^{-\gamma}/(2\sqrt{\pi}) \approx -0.1584$ and $-1/3\approx0.3333$). This numerical results thus point to the general validity of the proposed topological QED$_2$ as the continuum model  describing the role of SPT phases in lattice gauge  theories.

\begin{figure*}[t]
 \begin{centering}
  \includegraphics[width=2.1\columnwidth]{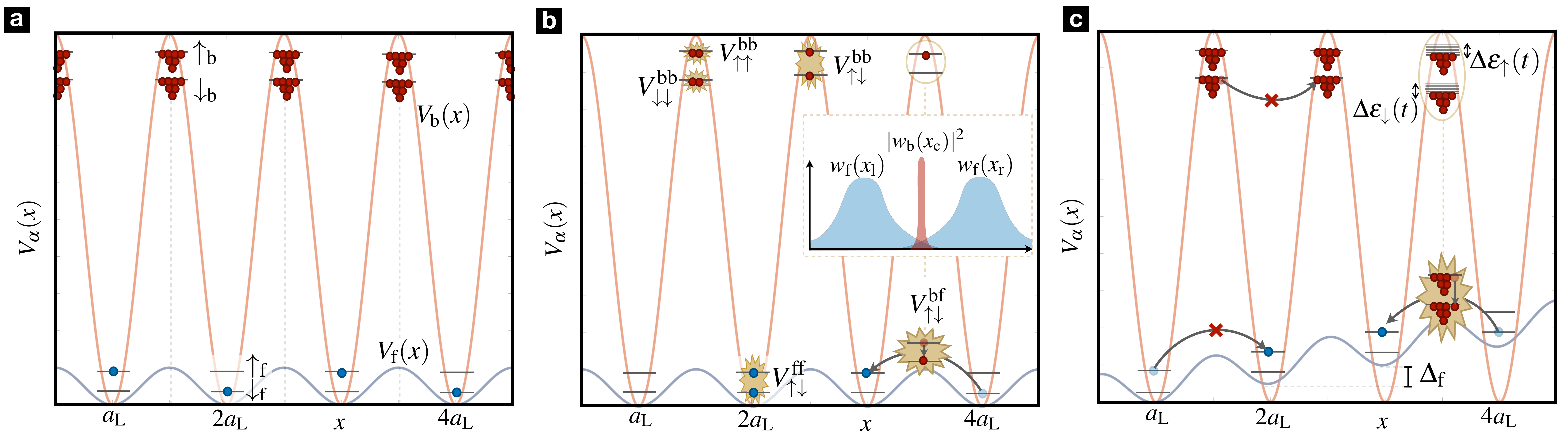}\\
  \caption{\label{Fig:cold_atoms_qs} {\bf Cold-atom QS of the topological Schwinger model:} {\bf (a)} Bose-Fermi mixture trapped in a 1D optical lattice formed by a red-detuned very deep lattice for the bosons (red circles trapped in two internal states at the maxima), and a blue-detuned shallower lattice for the fermions (blue circles trapped in two internal states at the minima). {\bf (b)} Dominant scattering processes, including on-site boson-boson and fermion-fermion Hubbard interactions, as well as boson-fermion spin-changing collisions allowed by the common overlap of neighboring Wannier functions for the fermions, weighted by the probability to find a boson on the intermediate link. {\bf (c)} Inhibition of the bosonic and fermionic bare tunnelings, either due to the very deep bosonic lattice, or to the application of a lattice tilting for the fermions. The spin-changing collisions can be assisted against the lattice tilting introducing a periodic modulation of the bosonic energy levels.}
\end{centering} 
\end{figure*}

\section{\bf Cold-atom quantum simulations of the topological Schwinger model}
\label{sec:cold_atoms_qs}
The remarkable level of isolation and control in several platforms of atomic, molecular and optical (AMO) physics, such as ultra-cold neutral atoms in periodic crystals made of light ~\cite{QS_cold_atoms} or trapped atomic ions in self-assembled Coulomb crystals~\cite{QS_trapped_ions}, has allowed to conduct experiments in the recent years bringing  the quantum simulation (QS) idea~\cite{feynman_qs} to a practical reality. Starting with the pioneering works showing that ultra-cold bosonic atoms can be used for the QS of the Bose-Hubbard model~\cite{analog_QS_bhm_exp}, a variety of schemes for the QS of condensed-matter models have been put forth~\cite{qs_book}. In recent years, several works have explored the possibility of using AMO systems for the QS of relativistic fermionic and bosonic quantum field theories~\cite{aqs_qft_fermions_2d,aqs_qft_fermions,aqs_qft_fermions_interactions,dqs_qft_phi4,O_N_models,aqs_phi_4},  theories of pure gauge fields~\cite{dqs_lattice_gauge_theories,aqs_qft_gauge_U(1)_qlinks,aqs_qft_gauge_U(1)_qlinks2,dqs_qft_SU(2)_qlinks, loops_and_strings_gauge_simulator}, theories for coupled Higgs and gauge fields~\cite{gauge_higgs_atoms,gauge_higgs_meurice}, and also theories of relativistic fermions interacting with Abelian and non-Abelian gauge fields~\cite{aqs_QED,aqs_qft_fermions_U(1)_qlinks,aqs_qft_fermions_SU(2)_qlinks}. Although most proposals have focused on ultracold atoms in optical lattices, we note that QS of gauge fields has also been explored for crystals of trapped ions, arrays of superconducting qubits and microwave resonators, and atom-ion mixtures~\cite{gauge_other_systems}.

Of particular relevance for our work will be the QS schemes for the simplest possible theory of matter coupled to gauge fields: quantum electrodynamics in (1+1) dimensions, or the so-called Schwinger model. We refer the reader to existing literature reviewing these schemes, and extensions thereof~\cite{qs_LGT_review,schwinger_mps_review}, and to the recent experiments~\cite{schwinger_experiments} realizing a digital QS (and a variational eigensolver of the massive Schwinger model). We note that there has also been recent progress on the analog QS of dynamical gauge fields~\cite{int_pat,density_dep_gauge_fields,z2_gauge_prop,z2_gauge_exp}. In the following, we will focus on analog QS of the topological Schwinger model~\eqref{eq:dimerised_KS_schwinger} using a Bose-Fermi mixture of ultra-cold neutral atoms trapped in an optical lattice.

\subsection{Bose-Fermi mixtures   for lattice gauge theories}

  We consider bosonic (fermionic) alkali atoms of mass $m_{\rm b}$ ($m_{\rm f}$), and select a pair of electronic states  $\{\ket{\uparrow_\alpha}=\ket{F_\alpha,M_{\alpha,\uparrow}},\ket{\downarrow_\alpha}=\ket{F_\alpha,M_{\alpha,\downarrow}}\}_{\alpha={\rm b,f}}$ from the hyperfine ground-state manifold, which will be referred to as spin components $\sigma=\uparrow,\downarrow$. The atoms are trapped in a spin-independent cubic optical lattice  formed by pairs of retro-reflected far-detuned laser beams with mutually-orthogonal linear polarizations, which lead   to  periodic ac-Stark shifts of strength $\{V_{0,j}^{\alpha}\}_{j=x,y,z}$~\cite{cold_atoms_ol}. The common  wavelength of the lasers  $\lambda_{\rm L}=2\pi/k_{\rm L}$ sets the lattice constant of the periodic light potential  to  $a_{\rm L}=\lambda_{\rm L}/2$ along each of the three axes. We will assume that $V_{0,x}^{\alpha}\ll V_{0,y}^{\alpha},V_{0,z}^{\alpha}$, such that the dynamics along the $y$ and $z$ axes is effectively frozen, and the  atomic mixture behaves as a 1D system  along the $x$ axis.

We consider  a regime  with a lattice depth that is much larger than the  atomic recoil energy $V_{0,x}^{\alpha}\gg E_{\rm R}^{\alpha}=k_{\rm L}^2/2m_\alpha$,  where we work with $\hbar=1$ as usual in the cold-atom literature. In this regime, the atoms are confined to small regions around the minima/maxima of the optical potential,  depending on the sign of the laser detuning. Therefore,  it is customary to introduce the so-called Wannier basis ~\cite{analog_QS_bhm,analog_QS_fhm,analog_QS_bfhm}, and define operators that create/annihilate bosonic $b^{\dagger}_{n,\sigma},b^{\phantom{\dagger}}_{n,\sigma'}$ and fermionic $f^{\dagger}_{n,\sigma},f^{\phantom{\dagger}}_{n,\sigma'}$ atoms localized at such positions, satisfying $[b^{\phantom{\dagger}}_{n,\sigma},b^{\dagger}_{m,\sigma}]=\{f^{\phantom{\dagger}}_{n,\sigma},f^{\dagger}_{m,\sigma}\}=\delta_{n,m}\delta_{\sigma,\sigma'}$. For a fermionic blue-detuned optical potential along the $x$ axis, the fermions would be trapped at the minima $x_n^{0}=na_{\rm L}$, where $n\in \mathbb{Z}_{N_{\rm s}}$ is the site index
and $N_{\rm s}$ is the number of lattice sites; whereas  the bosons are trapped at the maxima $x_n^{0}=(n+\half)a_{\rm L}$ of a red-detuned optical potential (see Fig.~\ref{Fig:cold_atoms_qs}{\bf (a)}).  In this way, one may  interpret that  fermions reside at the sites of a 1D chain and will be used to represent the matter sector of the topological Schwinger model, while bosons sit at the links and will be employed to simulate the   gauge field.

The dynamics of the Bose-Fermi mixture is controlled by the following lattice Hamiltonian
\beq
\label{eq:atom_tunneling}
\begin{split}
H_{\rm BF}&=\sum_{n,\sigma}(\epsilon^{\rm b}_{n,\sigma}b^{\dagger}_{n,\sigma}b^{\phantom{\dagger}}_{n,\sigma}-t_{\rm b}b^{\dagger}_{n,\sigma}b^{\phantom{\dagger}}_{n+1,\sigma}+{\rm H.c.})+\\
&+\sum_{n,\sigma}(\epsilon^{\rm f}_{n,\sigma}f^{\dagger}_{n,\sigma}f^{\phantom{\dagger}}_{n,\sigma}-t_{\rm f\hspace{0.5ex}}f^{\dagger}_{n,\sigma}f^{\phantom{\dagger}}_{n+1,\sigma}\hspace{0.5ex}+{\rm H.c.})+V_{\rm int},
\end{split}
\eeq
where we have introduced on-site energy terms $\epsilon^{\alpha}_{n,\sigma}$ that will depend on the energy of the electronic states including e.g. hyperfine energy of the electronic states  and an overall harmonic trapping potential. In this expression, we have introduced the tunneling strength $t_\alpha=4E_{\rm R}^\alpha ({V}_{0,x}^\alpha/E_{\rm R}^\alpha)^{3/4}{\rm exp}\{-2({V}_{0,x}^\alpha/E_{\rm R}^\alpha)^{1/2}\}/\sqrt{\pi}$~\cite{cold_atoms_ol}, which controls the hopping of fermionic/bosonic atoms between neighboring sites/links of the 1D chain. Additionally, at sufficiently low temperatures, the atoms  interact via $s$-wave scattering through a contact pseudo-potential~\cite{cold_atoms_ol}. In the Wannier basis, this yields various quartic terms  grouped in $V_{\rm int}$ as follows
\beq
\label{eq:atom_atom_interactions}
\begin{split}
V_{\rm int}=&\sum_{\boldsymbol{n},\boldsymbol{\sigma}}\half V^{\rm b,b}_{\boldsymbol{n},\boldsymbol{\sigma}}b^{\dagger}_{n_1,\sigma_1}b^{\dagger}_{n_2,\sigma_2}b^{\phantom{\dagger}}_{n_4,\sigma_4}b^{\phantom{\dagger}}_{n_3,\sigma_3}\\
+&\sum_{\boldsymbol{n},\boldsymbol{\sigma}}\half V^{\rm f,f}_{\boldsymbol{n},\boldsymbol{\sigma}}f^{\dagger}_{n_1,\sigma_1}f^{\dagger}_{n_2,\sigma_2}f^{\phantom{\dagger}}_{n_4,\sigma_4}f^{\phantom{\dagger}}_{n_3,\sigma_3}\\
+&\sum_{\boldsymbol{n},\boldsymbol{\sigma}} V^{\rm f,b}_{\boldsymbol{n},\boldsymbol{\sigma}}f^{\dagger}_{n_1,\sigma_1}b^{\dagger}_{n_2,\sigma_2}b^{\phantom{\dagger}}_{n_4,\sigma_4}f^{\phantom{\dagger}}_{n_3,\sigma_3}.\\
\end{split}
\eeq
Here, $\boldsymbol{n}=(n_1,n_2,n_3,n_4)$ and $\boldsymbol{\sigma}=(\sigma_1,\sigma_2,\sigma_3,\sigma_4)$ determine the positions and spin (i.e. electronic) states  of the colliding atoms $(n_1,\sigma_1)+(n_2,\sigma_2)\to (n_3,\sigma_3)+(n_4,\sigma_4)$, and $ V^{\alpha_1,\alpha_2}_{\hspace{0.5ex}\boldsymbol{n},\hspace{0.5ex}\boldsymbol{\sigma}}$ determines the strength of the corresponding scattering process, which involves a particular overlap of different Wannier functions. For deep optical lattices, these interaction strengths decay exponentially fast with the distance separating the corresponding Wannier centers, such that the leading contributions will be fermion-fermion (boson-boson) on-site (on-link) interactions, as well as boson-fermion  interactions between  site-link nearest neighbors (see Fig.~\ref{Fig:cold_atoms_qs}{\bf (b)}). 

As realized in~\cite{gauge_int_ang_mome_cons}, it is possible to exploit the conservation of total angular momentum in the $s$-wave atomic scattering  to identify those terms of Eq.~\eqref{eq:atom_atom_interactions} that  incorporate directly the gauge invariance of a LGT, such as the $U(1)$ gauge invariance of the standard  Schwinger model~\eqref{eq:standard_KS_schwinger}. In particular, with a suitable encoding, the boson-boson on-site interactions can give rise to the electric energy term $g^2L_n^2$. Additionally, the boson-fermion spin-changing scattering can lead to the gauge-invariant tunnelings $c_n^{\dagger} U^{\phantom{\dagger}}_n c_{n+1}^{\phantom{\dagger}}+c_{n+1}^{\dagger} U^\dagger_n c_{n}^{\phantom{\dagger}}$, where the link operators can be represented by spin ladder operators $U_n\to L_n^+,U^\dagger_n\to L_n^-$ satisfying  $[L^+_n,L^-_m]=2\delta_{n,m}L_n$, following the so-called quantum-link model approach to LGTs~\cite{q_link_models}. A detailed analysis of how this idea could be applied to a $^{23}$Na-$^{6}$Li Bose-Fermi mixture has been presented in~\cite{Schwinger_spin_changing_collisions_superlattice}, where the fermions were considered to be trapped in an optical superlattice providing a staggering of the on-site energies. This work showed that, in addition to the above mechanism to generate the gauge-invariant tunneling, the terms of Eqs.~\eqref{eq:atom_tunneling} and~\eqref{eq:atom_atom_interactions} that would violate gauge invariance can be neglected for  certain parameter regimes and initial states, as will be discussed below. Moreover,   it was shown that the additional spin-preserving  scattering between bosons and fermions can be expressed as a correction of the staggered fermion mass of the target Hamiltonian~\eqref{eq:standard_KS_schwinger}. In this way, there is a clear     route for the analog quantum simulation of the standard Schwinger model. Although challenging, we note that the  individual  required ingredients have all been shown experimentally, e.g. the types of optical lattices required, experiments with two-species Bose-Fermi mixtures, the control of Feshbach resonances, etc. The  future  technological quest is their combination  in a single experiment in the particular suitable regime  for a LGT simulator.

\subsection{Scheme  for the topological Schwinger model}

Let us now turn our attention to the cold-atom quantum simulation of the topological Schwinger model~\eqref{eq:dimerised_KS_schwinger}. Here, a natural idea would be to substitute the aforementioned staggered fermion superlattice by a superlattice where the distance between neighboring fermions varies within a two-site unit cell. Such optical potentials have already been realized experimentally,  and  yield a dimerized  tunneling that allows for the cold-atom quantum simulation~\cite{ssh_cold_atoms} of the SSH model of polyacetylene. One can check that this mechanism would also lead to a dimerization of the gauge-invariant tunneling~\eqref{eq:dimerised_KS_schwinger}, since the fermion-boson overlap of Wannier functions  determining the strength of the spin-changing collisions would also display a two-site unit cell periodicity. Unfortunately, the spin-conserving scattering terms also become inhomogeneous, and can no longer be simplified as  in the previous case~\cite{Schwinger_spin_changing_collisions_superlattice}, such that  the effective cold-atom Hamiltonian would contain additional terms that differ from the target model~\eqref{eq:dimerised_KS_schwinger}. Therefore, in this work, we introduce a scheme that avoids using superlattices, and achieves the desired dimerization by Floquet engineering through  a  periodic modulation on the bosonic degrees of freedom. We now detail the required ingredients. 

\subsubsection{Gauge sector and bosonic Hubbard interactions}

Let us start from the bosons and the gauge-field sector. First of all, we consider that the bosonic optical lattice is so deep $V_{0,x}^{\rm b}\gg E_{\rm R}^{\rm b}$ that the tunneling dynamics along the $x$-axis is also frozen   for the time-scales of interest $t_{\rm b}\propto E_{\rm R}^{\rm b}{\rm exp}\{-2({V}_{0,x}^\alpha/E_{\rm R}^{\rm b})^{1/2}\} \ll |V^{\alpha,\beta}_{\boldsymbol{n},\boldsymbol{\sigma}}|$. Using the Schwinger representation for spin operators in terms of bosons, referred to as rishons in the quantum link models,  one can encode the link gauge operators on even sites into the bosonic atoms as
\beq
\label{eq:even_Ls}
\begin{split}
L_{2n}&=\frac{1}{2}\left(b^{\dagger}_{2n,\uparrow}b^{\phantom{\dagger}}_{2n,\uparrow}-b^{\dagger}_{2n,\downarrow}b^{\phantom{\dagger}}_{2n,\downarrow}\right),\\
L_{2n}^+&=b^{\dagger}_{2n,\uparrow}b^{\phantom{\dagger}}_{2n,\downarrow}, \hspace{2ex}L_{2n}^-=b^{\dagger}_{2n,\downarrow}b^{\phantom{\dagger}}_{2n,\uparrow},
\end{split}
\eeq
and the link gauge operators on odd sites as
\beq
\label{eq:odd_Ls}
\begin{split}
L_{2n-1}&=\frac{1}{2}\left(b^{\dagger}_{2n-1,\downarrow}b^{\phantom{\dagger}}_{2n-1,\downarrow}-b^{\dagger}_{2n-1,\uparrow}b^{\phantom{\dagger}}_{2n-1,\uparrow}\right),\\
L_{2n-1}^+&=b^{\dagger}_{2n-1,\downarrow}b^{\phantom{\dagger}}_{2n-1,\uparrow}, \hspace{2ex}L_{2n-1}^-=b^{\dagger}_{2n-1,\uparrow}b^{\phantom{\dagger}}_{2n-1,\downarrow}.
\end{split}
\eeq
We note that the seemingly-arbitrary alternation in the definition of the link operators is important to obtain the gauge-invariant tunneling from the spin-changing collisions~\cite{gauge_int_ang_mome_cons,Schwinger_spin_changing_collisions_superlattice}. We also note that, since the bare bosonic tunneling can be neglected, the number of bosons with a particular spin $\sigma$ is a conserved quantity, and  can be used  to define the angular momentum of the link operators $\ell=N_{\uparrow}=N_{\downarrow}$ where, for simplicity, we assume that all sites are uniformly filled with bosons, and that the population is the same for both spin components. 

As announced above, the on-site boson-boson interactions of Eq.~\eqref{eq:atom_atom_interactions} can be expressed in terms of the electric-flux energy of the Schwinger model. For  a $^{23}$Na-$^{6}$Li Bose-Fermi mixture, and selecting the spin states $\ket{\uparrow_{\rm b}}=\ket{F_{\rm b}=1,M_{\rm b,\uparrow}=0},\ket{\downarrow_{\rm b}}=\ket{F_{\rm b}=1,M_{\rm b,\downarrow}=-1}$~\cite{Schwinger_spin_changing_collisions_superlattice}, the on-site Hubbard-type interactions can be expressed as follows
\beq
\label{eq:cold_atom_electric_energy}
V_{\rm int}^{\rm bb}=a\sum_{n=0}^{N_{\rm s}-1}\left(\delta_{\rm b}(-1)^nL_n+\frac{1}{2}g^2L_n^2\right).
\eeq
Here, we have introduced an effective gauge coupling $g$ and an effective lattice constant $a$, which can be expressed in terms of the atomic parameters
\beq
\label{eq:gauge_coupling_atoms}
g^2a=2\sqrt{\frac{8}\pi}k_{\rm L}\left(\frac{a^{\rm bb}_{0}-a^{\rm bb}_{2}}{6}\right)E_{\rm R}^{\rm b}\left(\frac{V_{0,x}^{\rm b}V_{0,y}^{\rm b}V_{0,z}^{\rm b}}{(E_{\rm R}^{\rm b})^3}\right)^{1/4}.
\eeq
Here, $a_{F_{\rm t}}^{\alpha\beta}$ stands for the $s$-wave scattering length between $\alpha$- and $\beta$-type atoms with total angular momentum $F_{\rm t}$. We note that, in addition to the desired electric-field energy, one gets in Eq.~\eqref{eq:cold_atom_electric_energy} an additional staggering $\delta_{\rm b}=(\ell-\half)g^2$ that must be carefully accounted for in the final effective Hamiltonian of the cold-atom mixture. 

\subsubsection{Matter sector and fermionic Hubbard interactions}

Let us now turn to the matter sector of the topological Schwinger model~\eqref{eq:dimerised_KS_schwinger}, which should be encoded into the fermionic atoms. As occurred for the bosons, the bare tunneling~\eqref{eq:atom_tunneling} must be inhibited as it does not preserve the gauge symmetry. However, in contrast to the bosonic case, we cannot simply consider a very deep lattice, since neighboring fermionic Wannier functions should overlap to allow for the desired gauge-invariant tunneling via the spin-changing collisions. A possible mechanism to inhibit  the bare tunneling could be  a staggered superlattice~\cite{Schwinger_spin_changing_collisions_superlattice}, but we have already advanced  that this would not suffice to achieve the desired topological Schwinger model~\eqref{eq:dimerised_KS_schwinger}. Therefore, we shall make use of a tilted optical lattice $\epsilon_{n,\sigma}^{\rm f}=\epsilon_{\sigma}^{\rm f}+\Delta_{\rm f} n$, which can be introduced by lattice acceleration~\cite{lattice_acceleration}, or magnetic-field gradients~\cite{b_field_gradient}. As a consequence of the tilting (see Fig.~\ref{Fig:cold_atoms_qs}{\bf (c)}), there is an energy penalty for the bare tunneling, such that fermions cannot hop between neighboring sites for the timescale of interest if $t_{\rm f}\ll\Delta_{\rm f}$, and the dynamics will be caused by  scattering.

 For  the particular $^{23}$Na-$^{6}$Li Bose-Fermi mixture, and selecting the spin states $\ket{\uparrow_{\rm f}}=\ket{F_{\rm f}=\half,M_{\rm f,\uparrow}=\half},\ket{\downarrow_{\rm f}}=\ket{F_{\rm f}=\half,M_{\rm f,\downarrow}=-\half}$, one can encode the discretized Dirac field into the fermion atomic operators of even and odd sites
\beq
\label{eq:odd_even_cs}
c_{2n}=\frac{1}{\sqrt{a}}f_{2n,\downarrow},\hspace{2ex} c_{2n-1}=\frac{1}{\sqrt{a}}f_{2n-1,\uparrow},
\eeq
and recover the desired algebra $\{c_n^{\phantom{\dagger}}, c_m^\dagger\}=\delta_{n,m}/a\to \delta(x-y)$  in the continuum limit. Moreover, the on-site fermion-fermion interactions of Eq.~\eqref{eq:atom_atom_interactions}, which  give the leading contribution of the fermion-fermion scattering, can be rewritten as  
\beq
\label{eq:cold_atom_fermion_fermion}
V_{\rm int}^{\rm ff}=a\sum_{n=0}^{N_{\rm s}-1}U\left(c^{\dagger}_{2n}c^{\phantom{\dagger}}_{2n}f^{\dagger}_{2n,\uparrow}f^{\phantom{\dagger}}_{2n,\uparrow}+c^{\dagger}_{2n-1}c^{\phantom{\dagger}}_{2n-1}f^{\dagger}_{2n-1,\downarrow}f^{\phantom{\dagger}}_{2n-1,\downarrow}\right),
\eeq
where we have introduced the Hubbard coupling strength
\beq
\label{eq:int_constant}
Ua=\sqrt{\frac{8}\pi}k_{\rm L}a^{\rm ff}_{0}E_{\rm R}^{\rm f}\left(\frac{V_{0,x}^{\rm f}V_{0,y}^{\rm f}V_{0,z}^{\rm f}}{(E_{\rm R}^{\rm f})^3}\right)^{1/4}.
\eeq
It follows from the above expression~\eqref{eq:cold_atom_fermion_fermion} that these interactions will have no effect if the state of the quantum simulator fulfills $f^{\dagger}_{2n,\uparrow}f^{\phantom{\dagger}}_{2n,\uparrow}\ket{\Psi(t)}=f^{\dagger}_{2n-1,\downarrow}f^{\phantom{\dagger}}_{2n-1,\downarrow}\ket{\Psi(t)}=0$ during the whole experiment. Since the bare tunneling is forbidden, and the lattice tilting forbids scattering terms that change the parity of fermions on each lattice site, it suffices to prepare an initial state $\ket{\Psi(0)}$  with no double occupancies (see the initial configuration in Fig.~\ref{Fig:cold_atoms_qs}{\bf (a)}), and the above condition to neglect these terms~\eqref{eq:cold_atom_fermion_fermion} will be fulfilled during the whole QS. 

\subsubsection{Gauge-matter coupling and  Floquet engineering of Bose-Fermi Hubbard interactions  }

Up to this point, the discussion is similar to the scheme in~\cite{Schwinger_spin_changing_collisions_superlattice}. In the following, we discuss the implementation of the dimerized gauge-invariant tunneling~\eqref{eq:dimerised_KS_schwinger}, where crucial differences arise. We now consider the fermion-boson site-link interactions~\eqref{eq:atom_atom_interactions}, which contain the  spin-changing collisions
\beq
\label{eq:spin_changing_scatt}
\begin{split}
V_{\rm int,1}^{\rm bf}&=\sum_{n=0}^{N_{\rm s}/2} V^{\rm f,b}_{\boldsymbol{n},\boldsymbol{\sigma}}f^{\dagger}_{2n-1,\uparrow}b^{\dagger}_{2n-1,\downarrow}b^{\phantom{\dagger}}_{2n-1,\uparrow}f^{\phantom{\dagger}}_{2n,\downarrow}\\
&+\sum_{n=0}^{N_{\rm s}/2}V^{\rm f,b}_{\boldsymbol{n},\boldsymbol{\sigma}}f^{\dagger}_{2n,\downarrow}b^{\dagger}_{2n,\uparrow}b^{\phantom{\dagger}}_{2n,\downarrow}f^{\phantom{\dagger}}_{2n+1,\uparrow}+{\rm H.c.}.
\end{split}
\eeq
These terms can be identified with the gauge-invariant tunneling $c_n^{\dagger} L^{+}_n c_{n+1}^{\phantom{\dagger}}+c_{n+1}^{\dagger} L^{-}_n c_{n}^{\phantom{\dagger}}$ using the atomic definitions of the link~\eqref{eq:even_Ls}-\eqref{eq:odd_Ls} and  fermion operators~\eqref{eq:odd_even_cs}. Note, however, that these terms will generally be inhibited by the tilting of the lattice $|V^{\rm f,b}_{\boldsymbol{n},\boldsymbol{\sigma}}|\ll\Delta_{\rm f}$, as occurred for the bare tunneling $t_{\rm f}\ll\Delta_{\rm f}$. 

The key idea of our proposed scheme is to assist these terms by introducing a time-periodic modulation of the bosonic on-site energies $\epsilon_{n,\sigma}^{\rm b}\to\epsilon_{n,\sigma}^{\rm b}(t)$ of Eq.~\eqref{eq:atom_tunneling}, which can be induced by a time-dependent magnetic field or by a weak ac-Stark shift with a time-dependent laser intensity (see Fig.~\ref{Fig:cold_atoms_qs}{\bf (c)}). In addition, we will adjust the  static part $\epsilon_{\sigma}^{\alpha}$ of the fermionic and bosonic on-site energies through an external constant magnetic field. Altogether, we are considering the  on-site energies
\beq
\label{eq:on-site_energies}
\epsilon_{n,\sigma}^{\rm b}(t)=\epsilon_{\sigma}^{\rm b}+\Delta^{\rm b}_{\sigma}\cos(\omega_{\rm d}t), \hspace{2ex}\epsilon_{n,\sigma}^{\rm f}=\epsilon_{\sigma}^{\rm f}+\Delta_{\rm f}n,
\eeq 
where $\epsilon_{\sigma}^{\alpha}=\epsilon_{F_\alpha}-g^{\rm L}_{F_\alpha}\mu_{\rm B}B_0M_{\alpha,\sigma}$ contains the hyperfine electronic energy $\epsilon_{F_\alpha}$, and the contribution from the linear Zeeman shift, where $g^{\rm L}_{F_\alpha}$ is the so-called Lande factor, $\mu_{\rm B}$ is Bohr's magneton, and $B_0$ the external constant magnetic field. In the simplest situation,  the periodic modulation stems from an ac-Stark shift $\Delta^{\rm b}_{\sigma}=-\alpha_\sigma(\omega_{\rm L})I_{\rm L}$,  where $\alpha_\sigma(\omega_{\rm L})$ is the spin-dependent dynamical polarizability of the state $\ket{F_{\alpha},M_{\alpha,\sigma}}$ irradiated by a far detuned laser beam of frequency $\omega_{\rm L}$ and intensity $I(t)=I_{\rm L}\cos(\omega_{\rm d}t)$, where $\omega_{\rm d}$ is the modulation frequency. 

We consider adjusting the lattice tilting,  external magnetic fields, and modulation frequency,  such that
\beq
\label{eq:modulation_parameters}
(\epsilon_{\uparrow}^{\rm b}-\epsilon_{\downarrow}^{\rm b})-(\epsilon_{\uparrow}^{\rm f}-\epsilon_{\downarrow}^{\rm f})=\Delta_{\rm f}+{\delta}_{\rm f}, \hspace{3ex}\omega_{\rm d}=2\Delta_{\rm f},
\eeq
where we have introduced a small energy mismatch $|\delta_{\rm f}|\ll\Delta_{\rm f}$. By writing the spin-changing scattering  in Eq.~\eqref{eq:spin_changing_scatt} in the interaction picture with respect to these on-site energy terms, one finds 
\beq
\label{eq:spin_chaing_t}
\begin{split}
V_{\rm int,1}^{\rm bf}(t)&=a\sum_{n=0}^{N_{\rm s}/2} V^{\rm f,b}_{\boldsymbol{n},\boldsymbol{\sigma}}c^{\dagger}_{2n-1}L^+_{2n-1}c^{\phantom{\dagger}}_{2n}\ee^{-\ii t({\delta}_{\rm f}+2\Delta_{\rm f})}F(t)\\
&+a\sum_{n=0}^{N_{\rm s}/2}V^{\rm f,b}_{\boldsymbol{n},\boldsymbol{\sigma}}c^{\dagger}_{2n}L^+_{2n}c^{\phantom{\dagger}}_{2n+1}\ee^{+\ii t{\delta}_{\rm f}}F^*(t)+{\rm H.c.},
\end{split}
\eeq
where we have introduced the periodic function $F(t)={\rm exp}\{\ii\eta_{\rm d}\sin(\omega_{\rm d}t)\}=\sum_{m\in\mathbb{Z}}J_{m}(\eta_{\rm d})\ee^{\ii m \omega_{\rm d}t}$, which is expressed in terms of Bessel functions of the first kind $J_m(\eta_{\rm d})$ with 
\beq
\label{eq:eta}
\eta_{\rm d}=(\Delta^{\rm b}_{\downarrow}-\Delta^{\rm b}_{\uparrow})/\omega_{\rm d}.
\eeq
Since we are working in the regime $|V^{\rm f,b}_{\boldsymbol{n},\boldsymbol{\sigma}}|\ll\Delta_{\rm f}$, and $\omega_{\rm d}=2\Delta_{\rm f}$,  the rapidly-oscillating terms  in the above expression~\eqref{eq:spin_chaing_t} can be neglected in a rotating-wave approximation. This is equivalent to the large-frequency limit of the so-called Floquet engineering based on periodic modulations~\cite{floquet_cold_atoms}. 

Moving back to a frame where the Hamiltonian is time-independent, we obtain the desired dimerized gauge-invariant tunneling   
\beq
\label{eq:eff_dimerised_gauge_tunneling}
V_{\rm int,1}^{\rm bf}=a\sum_n \left(m_{\rm s}(-1)^nc_n^{{\dagger}}c_n^{\phantom{\dagger}}+\left(\frac{1}{a}(1-\delta_n)c_n^{{\dagger}}L_{n}^{+}c_{n+1}^{\phantom{\dagger}}+{\rm H.c.}\right)\right),
\eeq
where we have introduced an additional  staggered mass $m_{\rm s}$, and the tunneling dimerization ${\delta}_{2n}=0$, and  ${\delta}_{2n-1}=\Delta$, with 
\beq
\label{eq:dimerisation_parameter}
m_{\rm s}=\delta_{\rm f},\hspace{2ex}\Delta=1-\frac{J_1(\eta_{\rm d})}{J_0(\eta_{\rm d})}.
\eeq 
Finally, the effective lattice constant of the topological Schwinger model~\eqref{eq:dimerised_KS_schwinger} is set by the overlap of the neighboring Wannier functions
\beq
\label{eq:lattice_constant_atoms}
\frac{1}{a}={\frac{k_{\rm L}\bigg(a^{\rm bf}_{\frac{3}{2}}-a^{\rm bf}_{\frac{1}{2}}\bigg)}{3\sqrt{2\pi}}}E_{\rm R}^{\rm f}\!\left(\!\!\frac{V_{0,x}^{\rm f}V_{0,x}^{\rm f}V_{0,x}^{\rm f}}{(E_{\rm R}^{\rm f})^3}\!\!\right)^{\!\!\fourth}\!\!\frac{m_{\rm f}}{\mu_{\rm bf}}\ee^{-\frac{\pi^2}{8}\sqrt{\frac{V_{0,x}^{\rm f}}{E_{\rm R}^{\rm f}}}}J_0(\eta_{\rm d}),
\eeq
where we have used the reduced mass $\mu_{\rm bf}=m_{\rm b}m_{\rm f}/(m_{\rm b}+m_{\rm f})$, and assumed that the bosons are subjected to a much tighter confinement than the fermions. 

It thus follows from our analysis that the effective dimerization $\Delta$ in the target model~\eqref{eq:dimerised_KS_schwinger} can be controlled by tuning the periodic-modulation parameters in Eq.~\eqref{eq:eta}. Likewise, the dimensionless gauge coupling $ga$ can be controlled by tuning the trapping and modulation parameters, as well as the various scattering lengths appearing in Eqs.~\eqref{eq:gauge_coupling_atoms} and~\eqref{eq:lattice_constant_atoms}. In this way, one could explore the full phase diagram of the topological Schwinger model provided that the additional staggered terms in Eqs.~\eqref{eq:cold_atom_electric_energy} and~\eqref{eq:eff_dimerised_gauge_tunneling} are carefully accounted for. We also note that the cold-atom simulator does not work in natural units, but leads instead to an effective speed of light $c=a_{\rm L}/a$. In any case, the  fields have the correct  energy dimensions $d_{c}=1/2$ and $d_{L}=0$, while the mass and gauge coupling have $d_{m_s}=d_g=1$, and the effective lattice constant $d_a=-1$.

\subsubsection{Suppression of additional spurious terms}

Let us note that, in addition to the terms leading to Eq.~\eqref{eq:eff_dimerised_gauge_tunneling}, there are other density-dependent tunneling terms inhibited in principle  by the tilting, which would also get activated by the periodic modulation. However, these terms can be neglected using the same argument used below Eq.~\eqref{eq:int_constant}. In addition,  there will be additional spin-conserving scattering terms that are not inhibited by the lattice tilting that require a more careful account. These terms can be expressed as 
\beq
\label{eq:spin_conserving_scatt}
\begin{split}
V_{\rm int,2}^{\rm bf}&=a\sum_{n,\sigma_2} V^{\rm f,b}_{\boldsymbol{n},\boldsymbol{\sigma}}c^{\dagger}_{2n}c^{\phantom{\dagger}}_{2n}(b^{\dagger}_{2n-1,\sigma_2}b^{\phantom{\dagger}}_{2n-1,\sigma_2}+b^{\dagger}_{2n,\sigma_2}b^{\phantom{\dagger}}_{2n,\sigma_2})\\
&+a\sum_{n,\sigma_2} V^{\rm f,b}_{\boldsymbol{n},\boldsymbol{\sigma}}c^{\dagger}_{2n+1}c^{\phantom{\dagger}}_{2n+1}(b^{\dagger}_{2n,\sigma_2}b^{\phantom{\dagger}}_{2n,\sigma_2}+b^{\dagger}_{2n+1,\sigma_2}b^{\phantom{\dagger}}_{2n+1,\sigma_2}).
\end{split}
\eeq
The fact that the fermionic density $c^{\dagger}_{n}c^{\phantom{\dagger}}_{n}$ is coupled equally to the rightmost and leftmost neighboring bosons, which underlies the form of Eq.~\eqref{eq:spin_conserving_scatt}, allows us to express this scattering  in terms of the finite gradient of the flux operators $(L_n-L_{n-1})$ via Eqs.~\eqref{eq:even_Ls}-\eqref{eq:odd_Ls}.  We note that this is precisely the condition that would not be satisfied had we used an optical superlattice, and which forced us to consider a Floquet-type alternative to achieve the tunneling dimerization. 

This expression~\eqref{eq:spin_conserving_scatt}
 can in turn be simplified further by applying Gauss' law $({L}_{n}-{L}_{n-1})=ac_{n}^{\dagger}c_{n} + \half((-1)^{n}-1)$ for the particular sector of the initial state of the cold-atom quantum simulator~\cite{Schwinger_spin_changing_collisions_superlattice}. This allows us to rewrite the spin-conserving scattering as a correction to the fermion staggered mass
\beq
\label{eq:spin_conserving_scatt_final}
V_{\rm int,2}^{\rm bf}=a\sum_n\tilde{m}_{\rm s}(-1)^nc_n^{\dagger}c_{n}^{\phantom{\dagger}},
\eeq
up to an irrelevant constant term, where we have introduced
\beq
\tilde{m}_{\rm s}={\frac{k_{\rm L}\bigg(a^{\rm bf}_{\frac{3}{2}}-a^{\rm bf}_{\frac{1}{2}}\bigg)}{\sqrt{\pi/2}}}\frac{\ell-2}{6}E_{\rm R}^{\rm f}\!\left(\!\!\frac{V_{0,x}^{\rm f}V_{0,x}^{\rm f}V_{0,x}^{\rm f}}{(E_{\rm R}^{\rm f})^3}\!\!\right)^{\!\!\fourth}\!\!\frac{m_{\rm f}}{\mu_{\rm bf}}\ee^{-\frac{\pi^2}{4}\sqrt{\frac{V_{0,x}^{\rm f}}{E_{\rm R}^{\rm f}}}}.
\eeq

As announced previously, the additional staggering terms must be carefully accounted for, as they do not appear in the target topological Schwinger model~\eqref{eq:dimerised_KS_schwinger}. In our case, their effect can be  cancelled by controlling the detuning of the periodic modulation~\eqref{eq:modulation_parameters}. It is straightforward to see that in the interaction picture with respect to $H_0=a\sum_{n}\left(\delta_{\rm b}(-1)^nL_n+(m_{\rm s}+\tilde{m}_{\rm s})c_n^{\dagger}c_{n}^{\phantom{\dagger}}\right)$, the  cold-atom Hamiltonian $H=V_{\rm int}^{\rm bb}+V_{\rm int,1}^{\rm bf}+V_{\rm int,2}^{\rm bf}=H_0+H_{t{\rm S}}$, leads to an effective time-independent Hamiltonian that coincides exactly with the topological Schwinger model $H_{\rm eff}=H_{t{\rm S}}$~\eqref{eq:dimerised_KS_schwinger}, provided that
\beq
\delta_{\rm f}=\half\delta_{\rm b}-\tilde{m}_{\rm s}.
\eeq
Accordingly, by tuning the energy mismatch between the energy-penalty and the modulation frequency~\eqref{eq:modulation_parameters}, it is possible to reach a situation where the staggered mass effectively vanishes from the Hamiltonian and, as desired, the Schwinger model only displays a topological mass stemming from the tunneling dimerization.
 
Let us finally note that, for simplicity of  exposition, we have assumed that the atoms are  subjected to a box trapping potential~\cite{box_potential_cold_atoms}, and thus  neglected additional inhomogeneities of the on-site energies~\eqref{eq:on-site_energies}. For other trapping potentials, it suffices to assume that they vary  slowly on the length-scale of the optical-lattice spacing, such that their presence will not compromise our assumptions and calculations above, and can be simply included as some additional inhomogeneities in the final effective Hamiltonian $H_{t{\rm S}}$. Alternatively, one could also decide to explore the more general phase diagram $(\Delta,m_{\rm s}a,ga)$, where the staggered mass $m_{\rm s}$ introduces another microscopic mechanism that competes with the dimerized topological mass $\Delta/a$, and will eventually lead to a trivial band insulator instead of the the SPT phase. In the future, it would be interesting to study the phase diagram of this model using the analytical and numerical techniques discussed in this work. We advance that, although the staggered mass will break the global symmetries~\cite{table_ti} that protect the SPT phase~\cite{spt_schwinger}, there will always be a local inversion symmetry, such that the SPT phase becomes a topological crystalline insulator~\cite{review_classification}.

\subsubsection{State preparation and readout}

Since  the model parameters $(ga,\Delta)$ are controlled  by atomic parameters in Eqs.~\eqref{eq:gauge_coupling_atoms},~\eqref{eq:lattice_constant_atoms} and~\eqref{eq:eta}, the cold-atom quantum simulator has the potential of realizing the interesting physics of topological QED$_2$ outlined in previous sections. In particular, by adiabatic state preparation, the cold-atom mixture could explore the full phase diagram of Fig.~\ref{Fig:qualitative_phase_diagram}, provided that some of the observables discussed in Sec.~\ref{sec:dmrg} can be measured experimentally. Density-type observables, such as the electric-field order parameter~\eqref{eq:magnifico_electric_parameter}, which is expressed in terms of the atomic densities~\eqref{eq:even_Ls}-\eqref{eq:odd_Ls}, are particularly promising as they are not modified by the various interaction pictures and rotating frames discussed above. We note that  the required spin-resolved density measurements can be performed  by optical imaging, either after a time-of-flight expansion or directly in situ~\cite{ketterle_reviews}. 
On the other hand, the topological correlator~\eqref{eq:magnifico_topological_order_quantity} will be modified by the tilting and effective staggering in the final rotating frame. In this case, it would be interesting to explore if  spin-resolved local measurements  by using recent quantum gas microscopes~\cite{microscopes} can give access to the local quantities, such that the corresponding rotating-frame order parameter can be constructed from a specific combination of the different local measurements. 

An alternative readout possibility would be to measure directly the edge excitations~\cite{edge_excitations_measurement}, which should display the dependence discussed in Ref.~\cite{spt_schwinger}. Although the above calculation uses Dirichlet boundary conditions, and would thus require  the use of box-like trapping potentials~\cite{box_potential_cold_atoms}, a similar behavior is expected for  confining potentials that increase at least quadratically with the boundary-to-center distance~\cite{edge_detection}. In that work, the authors also discuss how to use the Bragg signal due to Raman excitations to detect the presence of  localized zero-energy modes in a higher-dimensional setup, and these ideas could be adapted to the current topological Schwinger model~\eqref{eq:dimerised_KS_schwinger}. 

Finally, let us mention another readout strategy, which rests upon the possibility of measuring directly the underlying topological invariant. In the non-interacting regime,  we note that the Zak's phase~\cite{zak_phase,berry_rmp} has already been measured using Ramsey interferometry in  cold-atom experiments~\cite{ssh_cold_atoms}. As the gauge coupling is switched on, one expects that the fermionic excitations will become quasiparticles, which can be coupled to additional impurities in order to generalize the interferometric protocol to measure a many-body counterpart of the topological Zak's phase~\cite{many_body_zak}.

\section{\bf Conclusions and outlook}
\label{sec:concl}

In this work, we have explored the interplay of global and local symmetries, topology, and many-body effects in symmetry-protected topological phases of matter that arise naturally in lattice gauge theories. We have presented a detailed study of the topological Schwinger model by means of a class of $\mathbb{Z}_N$ lattice gauge theories showing that these models host a correlated topological phase for different values of  $N$, where interactions are mediated by the gauge field. By a careful finite-size scaling, we have shown that  this phase is stable in the large-$N$ limit, and that the phase boundaries are  in accordance to bosonization predictions of the  $U(1)$  topological Schwinger model. Finally, we have presented a detailed proposal for the realization of the topological Schwinger model exploiting spin-changing collisions in  boson-fermion mixtures of ultra-cold atoms loaded in a 1D optical lattice. We have shown that, by introducing a lattice tilting on the fermions and a periodic modulation of the bosons, the effective Hamiltonian coincides with the  target topological Schwinger model with parameters that can be controlled by tuning the microscopic cold-atom parameters.
In combination with the recent work~\cite{spt_gauge}, which studies the appearance of SPT phase in quantum link ladders, our work opens an interesting route to study topological phases of matter in gauge theories, either using some of the theoretical tools hereby developed, or via cold-atom experiments using the  scheme proposed in this work.  Hopefully, these results will stimulate further work in this subject, exploring interesting questions such as the interplay of topological features with non-perturbative  effects in LGTs, such as screening, confinement, and string-breaking. In addition, it will be very interesting to explore the appearance of topological properties in a non-equilibrium situation where the quantum-mechanical $\hat{\theta}$ angle is quenched. For a sudden quench in the classical vacuum $\theta$ angle, very interesting dynamical topological phase transitions have been identified in~\cite{dynamical_topo_transition_theta_angle} by exploring bulk phenomena. Our approach would allow to have an edge perspective of these phenomena, and explore the consequences of a quantum-mechanical $\hat{\theta}$ angle.

\acknowledgements
E.E. and G.M. are partially supported through the project "QUANTUM" by Istituto Nazionale di Fisica Nucleare (INFN) and through the project "ALMAIDEA" by University of Bologna. S.P.K. acknowledges support from the STFC grant ST/P00055X/1. A.B. acknowledges support from the Ram\'on y Cajal program RYC-2016-20066, MINECO project FIS2015-70856-P, and CAM/FEDER Project S2018/TCS-4342 (QUITEMAD-CM).

 %%%%%%%%%%%%%%%%%%%%%%%%%%%%%%%%%%%%%%%%%%%
%\bibliographystyle{apsrev4-1}
%\bibliography{biblio_intro_qed2,biblio_schwinger}
%\end{document}

\end{document}